# Spin-Mediated Consciousness: Theory, Experimental Studies, Further Development & Related Topics

Huping Hu ∗ & Maoxin Wu

(Dated: November 7, 2007)

Biophysics Consulting Group, 25 Lubber Street, Stony Brook, New York 11790, USA

## ABSTRACT

A novel theory of consciousness is proposed in this paper. We postulate that consciousness is intrinsically connected to quantum spin since the latter is the origin of quantum effects in both Bohm and Hestenes quantum formulism and a fundamental quantum process associated with the structure of space-time. That is, spin is the "mind-pixel." The unity of mind is presumably achieved by entanglement of the mind-pixels. Applying these ideas to the particular structures and dynamics of the brain, we have developed a detailed model of quantum consciousness. Experimentally, we have contemplated from the perspective of our theory on the possibility of entangling the quantum entities inside the brain with those in an external anaesthetic sample and carried out experiments toward that end. We found that applying magnetic pulses to the brain when a general anaesthetic sample was placed in between caused the brain to feel the effect of said anaesthetic for several hours after the treatment as if the test subject had actually inhaled the same. The said effect is consistently reproducible on all subjects tested. We further found that drinking water exposed to magnetic pulses, laser light, microwave or even flashlight when an anaesthetic sample was placed in between also causes consistently reproducible brain effects in various degrees. Further, through additional experiments we have verified that the said brain effect is the consequence of quantum entanglement between quantum entities inside the brain and those of the chemical substance under study induced by the photons of the magnetic pulses or applied lights. With the aids of high-precision instruments, we have more recently studied non-local effects in simple physics system such as a liquid. We have found that the pH value, temperature and gravity of a liquid such as water in the detecting reservoirs can be non-locally affected through manipulating a liquid in a remote reservoir quantum-entangled with the former. In particular, the pH value changes in the same direction as that being manipulated; the temperature can change against that of local environment; and the gravity can change against local gravity. Again, these non-local effects are all reproducible and have practical applications in many areas. We suggest that they are mediated by quantum entanglement between nuclear and/or electron spins in treated water and discuss the profound implications of these results. To be more comprehensive, this paper now also includes, as appendices, materials on further development of the theory and related topics.

---

∗ To whom correspondence should be addressed. E-mail: hupinghu@quantumbrain.org

# TABLE OF CONTENT

# Part I. Spin-Mediated Consciousness Theory: Possible Roles of Neural Membrane Nuclear Spin Ensembles and Paramagnetic Oxygen

## 1. INTRODUCTION

Experimentally, tremendous progress has been made in neuroscience over the last several decades. Theoretically, numerous versions of quantum and non-quantum consciousness theories have been proposed over the recent years (1-8). But, at this stage almost all these theories are speculative and none is commonly accepted. Philosophically, the age-old debate about consciousness has intensified like a raging fire (9-11). However, despite all these efforts, what is and causes consciousness remains a deep mystery. In this paper, we propose a novel theory of consciousness with the hope that it would shed some light on these issues.

As further discussed below, spin is a very fundamental quantum process associated with the structure of space-time (12-14). Indeed, modern physics leads us right down to the microscopic domain of space-time where various models of elementary particles and even space-time itself are built with spinors (15). On the other hand, neural membranes are saturated with spin-carrying nuclei such as $^1$H, $^{13}$C and $^{31}$P. Indeed, both MRI and fMRI are based on the abundance of $^1$H in human body. Neural membranes are matrices of brain electrical activities and play vital roles in the normal functions of a conscious brain and their major molecular components are phospholipids, proteins and cholesterols. Each phospholipid contains 1 $^{31}$P, 1.8% $^{13}$C and over 60 $^1$H its lipid chains. Similarly, neural membrane proteins such as ion channels and neural transmitter receptors also contain large clusters of spin-carrying nuclei. Therefore, we strongly believe that Nature has utilized quantum spin in constructing a conscious mind.

Very importantly, we believe that the mechanism of anesthetic action is closely related to the inner workings of consciousness. But how general anesthetics work is itself a 150-year old mystery (16, 17). We have already proposed within the framework of conventional neuroscience that anesthetic perturbations of oxygen pathways in both neural membranes and proteins are possibly involved in general anesthesia (17). Each $O_2$ contains two unpaired valence electrons thus is strongly paramagnetic and at the same time chemically reactive as a bi-radical. It is capable of producing a large fluctuating magnetic field along its diffusing pathway thus serves as a natural contrast agent in MRI (18). The existence of unpaired electrons in stable molecules is very rare indeed. $O_2$ are the only paramagnetic specie to be found in large quantities in the brain besides enzyme-produced nitric oxide (NO). In addition, $O_2$ is an essential component for energy production in the central nervous system.

Both $O_2$ and NO, the latter being a unstable free radical with one unpaired electron and a recently discovered small neural transmitter, are well known in spin chemistry - a field focused on the study of free-radical mediated chemical reactions where very small magnetic energies can change non-equilibrium spin conversion process (19,20). Thus $O_2$ and NO may serve as spin-catalysts in consciousness-related neural biochemical reactions such as those dual path reactions initiated/driven by free radicals (21).

## 2. NATURE OF SPIN

Unlike mass and charge that enter a dynamic equation as arbitrary parameters, spin reveals itself through the structure of the relativistic quantum equation for fermions such as electrons (12). Penrose had considered early on that spin might be more fundamental than space-time and invented spinor and twistor algebras for a combinatorial description of space-time geometry (13, 14). Bohm and Hiley generalized the twistor idea to Clifford algebra as a possible basis for describing Bohm's implicit order (22). Recently various spin foams have been formulated as extensions to Penrose's spin networks for the purpose of constructing a consistent theory of quantum gravity (23, 24).

In Hestenes' geometric picture, the zitterbewegung associated with the spin of the Dirac electron is qualitatively shown to be responsible for all known quantum effects of said electron and the imagery number *i* in the Dirac equation is said to be due to electronic spin (25). Second, in Bohmian mechanics the quantum potential is responsible for quantum effects (26). Salesi and Recami has recently shown that said potential is a pure consequence of "internal motion" associated with spin evidencing that the quantum behavior is a direct consequence of the fundamental existence of spin (27). Esposito has expanded this result by showing that "internal motion" is due to the spin of the particle, whatever its value (28). Very recently, Bogan has further expanded these results by deriving a spin-dependent gauge transformation between the Hamilton-Jacobi equation of classical mechanics and the time-dependent Shrődinger equation of quantum mechanics that is a function of the quantum potential in Bohmian mechanics (29). Third, Kiehn has shown that the absolute square of the wave function could be interpreted as the vorticity distribution of a viscous compressible fluid that also indicates that spin is the process driving quantum effects (30).

Many others have also study the nature of spin from both classical and quantum-mechanical perspectives (31-33). For example, Newman showed that spin might have a classical geometric origin. Galiautdinov has considered a theory of spacetime quanta and suggested that spin might manifest the atomic structure of space-time (32). Finkelstein is proposing that spin derives from a swap – a projective permutation operator, and the two-valued spin representation from a deeper 2-valued statistics (33). Sidharth has showed that spin is symptomatic of the non-commutative geometry of space-time at the Compton scale of a fermion and the three dimensionality of the space result from the spinorial behavior of fermions (34-35). He further showed that mathematically an imaginary shift of the spacetime coordinate in the Compton scale of a fermion introduces spin ½ into general relativity and curvature to the fermion theory (34-35). The reason why an imaginary shift is associated with spin is to be found in the quantum mechanical zitterbewegung within the Compton scale and the consequent quantized fractal space-time (34-35). Further, according to Sidharth, a fermion is like a Kerr-Newman Black Hole within the Compton scale of which causality and locality fails (34-35).

**3. MYSTERY OF GENERAL ANESTHESIA**

There is no commonly accepted theory on how general anesthetics work (16,17). However, there are two schools of thoughts on the issue. The first and oldest is the "lipid theory" which proposes that anesthetics dissolve into cell membranes and produce common structural perturbation resulting in depressed function of ion channels and receptors that are involved in brain functions (16). The second, more popular and recent theory is the "protein theory" which suggests that anesthetics directly interact with membrane proteins such as ion channels and receptors that are involved in brain functions. But the protein theory doesn't seem to square well with the low affinity and diversity of the general anesthetics. There is no direct experimental evidence to support either theory (17).

However, both theoretical and experimental studies have shown that many general anesthetics cause changes in membrane structures and properties at or just above the clinical concentrations required for anesthesia (16,36-37). Since both $O_2$ and general anesthetics are hydrophobic, we have proposed within the framework of conventional neuroscience that general anesthetic may cause unconsciousness by perturbing $O_2$ pathway in neural membranes and $O_2$-utilizing proteins, such that the availability of $O_2$ to its sites of utilization is reduced, which in turn triggers cascading cellular responses through $O_2$-sensing mechanisms, resulting in general anesthesia (17).

We have also been asking the question whether anesthetic perturbations of neural membranes and oxygen pathways themselves are the direct cause of unconsciousness. This conjuncture requires that $O_2$ and neural membranes be directly involved in consciousness. Indeed, the low affinity, diversity and pervasiveness of general anesthetics point us to this direction. If we assume that consciousness is an emergent property of the brain (9) and further liken consciousness to the formation of ice at $0\ ^oC$, the anesthetic action would be like the action of salt which prevents ice formation.

**4. NATURE OF CONSCIOSNESS**

There is no coherent view as to what is and causes consciousness. Some neuroscientists would say that it is the connections between the neurons and the coherent firing patterns thereof (2, 6). Some physicists would propose that it is connected to the measurement problem in quantum theory and thus the solution lies there (1,3,5,7). A few philosophers would suggest that it is an emergent property of the complex brain (9) or a new kind of properties and laws are required (11). For sure such disarray has its historical reasons. Ever since Descartes promoted his dualism philosophy in the 17$^{th}$ Century, science has been for the most part steered clear from this subject until very recently.

Philosophically, Searle argues that consciousness is an emergent biological phenomenon thus cannot be reduced to physical states in the brain (9). Chalmers argues that consciousness cannot be explained through reduction, because mind does not belong to the realm of matter (11). In order to develop a consciousness theory based on this approach, Chalmers suggests expanding science in a way still compatible with today's scientific knowledge and outlines a set of fundamental and irreducible properties to be added to space-time, mass, charge, spin etc. and a set of laws to be added to the laws of Nature (11). Further, he considers that information is the key to link consciousness and the physical world.

On the theoretical front, there are quite a few quantum theories of mind (1,3-5,7,8). Among these, Penrose's Objective Reduction ("OR") together with Hameroff's microtubule computation is perhaps the most popular, and the combination of the two produced the Orchestrated Objective Reduction ("Orch OR") in microtubules (1,7,8). According to Penrose, each quantum state has its own space-time geometry, thus superposition of quantum states entails superposition of different space-time geometries (1,7). Under certain conditions, such space-time geometric superposition would separate under its own "weight" through a non-computable process, which in turn would collapse said quantum state superposition (1,7). Hameroff suggested that such self-organized OR could occur in microtubules because of their particular structures, thus, born the Orch OR (8). According Orch OR, each collapse of macroscopic space-time geometry superposition corresponds to a discrete conscious event (8). In addition, it seems that Penrose accepts a separate mental world with grounding in the physical world (1,7).

There are also a number of theories based on conventional neuroscience (2,6). Our view on these is that whatever the final accepted version based on neuroscience ("classical physics"), it could be accepted as classically correct. The reason is that we must rely on the classical parts of the brain working according to conventional neuroscience to provide us the necessary neural components and wirings such as coherent neural firings, neurotransmitter releases and neural plasticity to support any realistic quantum activities of the brain. The situation is much like that in quantum computation where classical components form the supporting system of a quantum computer. Without these classical components, quantum computation could not be implemented at all.

In comparison, our working philosophy is that consciousness is grounded at the bottom of physical reality and emerges from the collective dynamics of known physical candidates inside the brain. Next we ask "where" and "how." To answer these, we take the reductionist approach both down to the end of physics to see what is left there and to the microscopic domain of a neuron to see what may be really important for the functioning of a conscious brain. What we found is that there is almost nothing left at the end of physics except the fundamental ideas of quantized space-time and spin. On the other hand, we found that what may be really important in the microscopic domain of a neuron are the nuclear spin ensembles and the fluxes of biologically available unpaired electrons carried by small molecules such as $O_2$ and NO. Naturally, we draw the conclusion that quantum spin together with its connection to space-time geometry is needed to ground consciousness in physical reality such that conscious experience emerges from the successive collapses of various entangled neural nuclear spin states.

Specifically, we try to answer these questions: (1) what are the neural substrates of consciousness, (2) what physical processes are involved in conscious experience, (3) what physical and biochemical process are involved in connecting consciousness to the classical neural networks of the brain and, (4) what binding mechanism allows the mind to achieve unity.

## 5. GENERAL POSTULATES

With above discussions in mind, we present the following Postulates: (a) Consciousness is intrinsically connected to quantum spin; (b) The mind-pixels of the brain are comprised of the nuclear spins distributed in the neural membranes and proteins, the pixel-activating agents are comprised of biologically available paramagnetic species such as $O_2$ and NO, and the neural memories are comprised of all possible entangled quantum states of the mind-pixels; (c) Action potential modulations of nuclear spin interactions input information to the mind pixels and spin chemistry is the output circuit to classical neural activities; and (d) Consciousness emerges from the collapses of those entangled quantum states which are able to survive decoherence, said collapses are contextual, irreversible and non-computable and the unity of consciousness is achieved through quantum entanglement of the mind-pixels.

In Postulate (a), the relationship between quantum spin and consciousness are defined based on the fact that spin is the origin of quantum effects in both Bohm and Hestenes quantum formulism (25, 27-29) and a fundamental quantum process associated with the structure of space-time (12-14). Combining this fundamental idea with those stated in Postulates (b), (c) and (d) allows us to build a qualitatively detailed working model of consciousness as discussed later.

In Postulate (b), we specify that the nuclear spins in both neural membranes and neural proteins serve as the mind-pixels and propose that biologically available paramagnetic species such as $O_2$ and NO are the mind-pixel activating agents. We also propose that neural memories are comprised of all possible entangled quantum states of mind-pixels. This concept of memory is an extension to the associative memory in neuroscience as will be discussed later.

In Postulate (c), we propose the input and output circuits for the mind-pixels. As shown in a separate paper, the strength and anisotropies of nuclear spin interactions through J-couplings and dipolar couplings are modulated by action potentials. Thus, the neural spike trains can directly input information into the mind-pixels made of neural membrane nuclear spins. Further, spin chemistry can serve as the bridge to the classical neural activity since biochemical reactions mediated by free radicals are very sensitive to small changes of magnetic energies as mentioned earlier and further discussed later (19-21).

In Postulate (d), we propose how conscious experience emerges. Since there are several interpretations of the measurement problem in quantum mechanics, we choose to accept the collapsing view (1,7). Thus, we adopt a quantum state collapsing scheme from which conscious experience emerges as a set of collapses of the decoherence-resistant entangled quantum states. We further theorize that the unity of consciousness is achieved through quantum entanglements of these mind-pixels (5).

## 6. ILLUSTRATIONS OF SPIN-MEDIATED CONSCIOUSNESS

### Overview

Figure 1 is a highly schematic drawing of the overall picture of a spin-mediated consciousness model proposed herein. At the top of Figure 1, a two-neuron network is shown. The connections are self-explanatory. The neural activities of the postsynaptic membrane are immediately shown below the neurons in Figure 1. These activities include biochemical reactions immediately following the release of neurotransmitters into the synaptic cleft, the ensuing collective activities of multiple ion channels and the action potentials and their propagations thereof, and other enzymatic activities.

The present model is mainly concerned with the dynamics of the nuclear spin ensembles in neural membranes and proteins such as those on the dendrites and soma under modulations by action potentials and activations by paramagnetic species such as rapidly tumbling and diffusing $O_2$ and neural transmitter NO, and the connections of such dynamics to conscious experience. The input and out interface of the neural nuclear spin ensembles are schematically shown in the middle of Figure 1. On the bottom of Figure 1, what the conscious brain perceives is schematically shown. The neural substrates and mechanism of the spin-mediated consciousness are described below.

### Nuclear spin ensembles

Figure 2 shows side-by-side a typical phospholipid found in neural membranes together with the diffusing $O_2$. A similar but much complex picture can be drawn for a neural protein. As can be seen, each phospholipid molecule contains 1 $^{31}P$, less than 1 $^{13}C$ (1.1%) and more than 60 $^1H$ on its two lipid chains. Because the small mobility of nuclei and weak interactions of nuclear spins with their environments, nuclear spins have long relaxation time after excitations (18). This property of nuclear spins is ideal for them to serve as the mind-pixels. Importantly, these nuclear spins can form various intra/inter-molecularly entangled quantum states under different external activations through J-couplings and dipolar coupling (38-40).

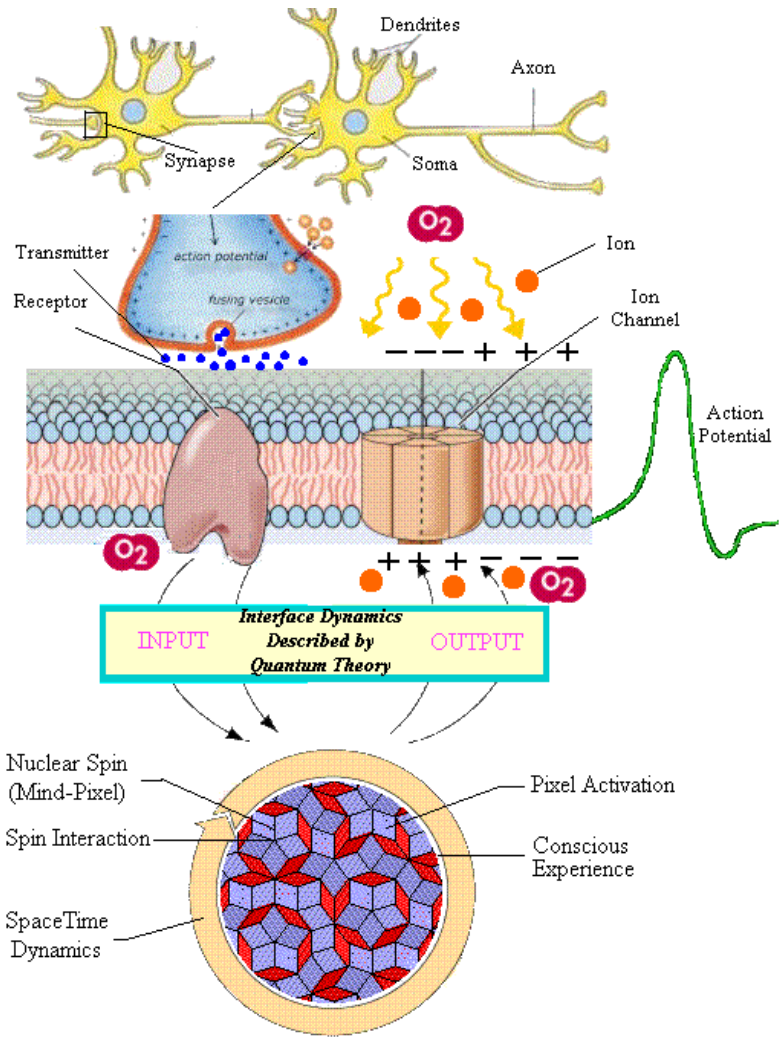

Figure 1. Illustration of spin-mediated consciousness theory. The drawing is self-explanatory except the part dealing with conscious experience See text for detailed explanations.

**Activation Agents**

The unpaired electrons attached to the paramagnetic species such as $O_2$ and NO can interact with nuclear spins through their large magnetic dipoles and collision-induced Fermi-contact mechanism (41) thus activating the neural nuclear spin ensembles. Indeed, because the magnetic dipole moment of an unpaired electron is 658 times larger than that of the $^1H$ nucleus, $O_2$ and NO can respectively produce magnetic fields 1,316 and 658 times larger than $^1H$ (41). In addition, $O_2$ and NO are hydrophobic small molecules so their concentrations in neural membranes are much higher than in aqueous solutions such as cytoplasma (42). Thus, as they rapidly tumble and diffuse, they produce microscopically strong and fluctuating magnetic fields. $O_2$ are the predominant sources of internal magnetic fields in neural membranes as evidenced by the strong effect of $O_2$ on spin-spin and spin-lattice relaxation rates (42,43).

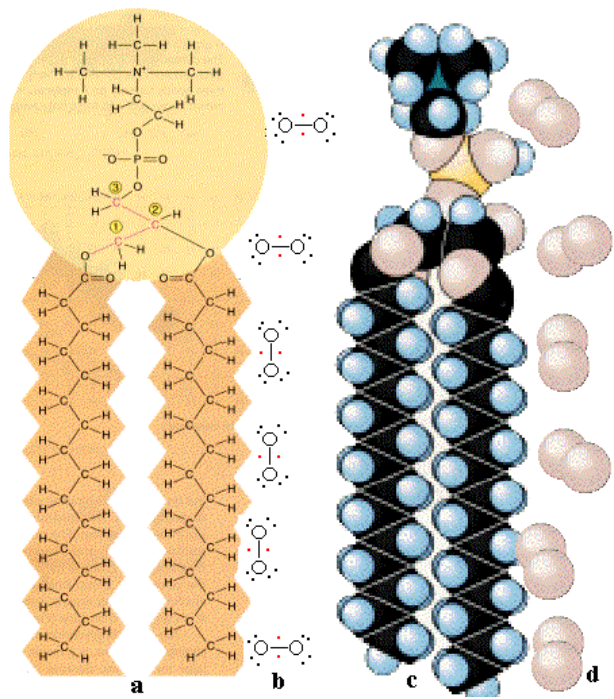

Figure 2. Schematic drawings of chemical structure (a) and atomic model (c) of a typical phospholipid with diffusing $O_2$ in Lewis structure (b) and atomic model (d) shown along the side. Two unpaired electronic electrons of $O_2$ are shown in Lewis structure as two red dots (b). In the atomic model, the white, black, purple and orange balls respectively represent hydrogen, carbon, oxygen and phosphonium atoms. $^1H$, $^{13}C$ and $^{31}P$ are spin-carrying nuclei.

### Spin-mediated mechanism

The mechanism of spin-mediated consciousness is concisely stated here and the related issues such as decoherence effect are treated in next subsection. Through action potential modulated nuclear spin interactions and paramagnetic $O_2$/NO driven activations the nuclear spins inside neural membranes and proteins form various entangled spin states some of which survive rapid decoherence through quantum Zeno effects or in decoherence-free subspaces and then collapse contextually through non-computabe and irreversible means thus producing consciousness and, in turn, the collective spin dynamics associated with said collapses have effects through spin chemistry on classical neural activities thus influencing the classical neural networks of the brain. Our proposal calls for extension of associative encoding of neural memories to the dynamical structures of neural membranes and proteins.

Therefore, according to the present theory, (a) the dynamical nuclear spin ensembles are the "mind-screen" with nuclear spins as its pixels, (b) the neural membranes and proteins are the mind-screen and memory matrices; and (c) the fluxes of biologically available paramagnetic $O_2$/NO are the beam for pixel-activations. Together, they form the neural substrates of consciousness. An analogy to this is the mechanism of liquid crystal display (LCD) where information-carrying electric voltages applied to the pixel cells change the optical properties of the constituent molecules such that when lights pass through these cells their phases get rotated differently which in turn represent different information to the viewer of the LCD screen (44).

**Quantum coherence and decoherence**

The decoherence effect which causes a quantum system to lose quantum coherence through interactions with its environment is a major concern for any quantum theory of the brain and is hotly debated (45-46). However, nuclear spins only have weak interactions with their environments thus long relaxation time after excitation (18). Indeed, there are both theoretical and experimental studies (38-40, 47-50) indicating the possibility of large-scale quantum coherence with entanglement in the nuclear spin ensembles distributed in the neural membranes and proteins. Paradoxically, the interactions of the neural membrane nuclear spin ensembles with their noisy brain environments may even enhance quantum coherence through quantum Zeno effect which prevents a quantum system to evolve/decohere through repeated collisions with their environments (5). Further, studies show that decoherence-free subspaces can exist within the Hilbert space of a complex quantum system (51).

In a series of experiments, Khitrin *et al* have demonstrated that a cluster of dipolar coupled to $^1$H nuclear spins in the molecules of a nematic liquid crystal at room temperature can be manipulated to achieve long-lived intra-molecular quantum coherence with entanglement such that a large amount of information may be stored in said cluster (38-40). In particular, they have succeeded in storing at room temperature a 2D pattern consisting of 1025 bits of information in the proton nuclear spin states of a molecular system of said nematic liquid crystal and then retrieved the same as a stack of NMR spectra (38-40).

Second, about a decade ago Warren *et al* discovered long-ranged intermolecular multiple-quantum coherence in NMR spectroscopy and imaging and have since successfully applied said coherence as MRI contrast agents (47-48). Indeed, they found that even the intermolecular dipolar couplings of the nuclear spins at distances larger than 10 microns are not averaged away by diffusions (47-48).

Third, Julsgaard *et al* have first theoretically predicted and then experimentally demonstrated at room temperature a long-lived entanglement of two macroscopic spin ensembles formed by two caesium gas samples each of which contains about $10^{12}$ atoms (49). The entangled spin-state can be maintained for 0.5 milliseconds and was generated via interactions of the samples with a pulse of light (49). The state they demonstrated is not a maximally entangled "Schrödinger cat" state but a state similar to a two-mode squeezed state; thus, it is an example of a non-maximally entangled state (49). In addition, Kun *et al* have theoretically predicted a "Schrödinger cat" state to be found in highly-excited and strongly-interacting many-body system (50).

These results apparently contradict the claim that there is no large-scale quantum coherence in the noisy brain (45). At least, this claim does not seem to apply to the nuclear spin ensembles in neural membranes and proteins. Further, in the case of Penrose-Hameroff microtubule model (8), it was strongly argued by Hagan *et al* that Tegmark's theoretical calculations also do not apply there (46). Indeed, even if we assume this claim is true, consciousness can still emerge from the statistical mixtures of coherent and incoherent quantum states of the mind pixels, as long as we either accept some kind of emergence theory (9) or take a dualistic approach (4,11) as will be discussed again later.

**"Consciousness explained"**

The explanation of consciousness in accordance with our spin-mediated theory is schematically shown at the bottom of Figure 1. The geometry inside the spinning circle represents conscious experience and is part of a Penrose tiling (1). It symbolizes that consciousness emerges from the non-computable collapses of entangled quantum states of the mind-pixels under the influence of spacetime dynamics schematically shown as the spinning circle. The edges in the Penrose tiling represent the nuclear spins in neural membranes and proteins as mind-pixels, the nodes represent interactions between these nuclear spins through J-couplings and dipolar couplings and the colors represent activations of mind-pixels by biologically available paramagnetic species such as $O_2$ and NO. The whole tiling pattern in Figure 1 represents conscious experience and the underlying spacetime geometry. This pattern successively evolves under repeated activations by the paramagnetic species representing successive collapses of the entangled quantum states of the mind-pixels that have survived decoherence as a stream of conscious experience.

We adopt Penrose's long-standing view that human thought may involve non-computable processes, as Gödel's theorem of incompleteness would suggest (1,7). According to Gödel, any consistent system of axioms beyond a certain basic level of complexity yields statements that cannot be proved or disproved with said axioms. Yet human can decide whether those statements are true, thus human thought cannot be reduced to a set of rules or computations (1,7). So where can one find non-computable process in physics? Obviously it cannot be found in classical physics because classical physics is deterministic so, in principle, can be simulated by a computer (1,7). Thus, Penrose reasoned that some kind of non-computable quantum process must be involved in consciousness and further suggested gravity-induced reduction ("R") process of quantum state superposition to be the candidate (1,7). One may recall that, according to Einstein's theory of general relativity, gravity is space-time geometry and, further, as we have discussed before quantum mechanical spin is associated with the structure of space-time. Therefore, the quantum state of spin must be connected to the underlying space-time geometry. However, we still have the task of working out the details in future research. This will be especially difficult because at the present we do not have a satisfactory theory of quantum gravity.

**Alternative approaches to consciousness**

If we assume that there is no large-scale quantum coherence in the noisy brain because of decoherence (45), how can consciousness still emerge from the statistically mixed quantum states of the nuclear spin ensembles in neural membranes and proteins? There are indeed at least two ways out. The first is to adopt an emergence theory (9) and the second is to take a dualistic approach (4, 11). Here, we will focus our discussion on the dualistic approach.

In such approach we can propose that mind has its own independent existence and reside in a pre-space-time domain. Then, the question becomes how does mind process and harness the information from the brain so that it can have conscious experience? We can theorize that conscious experience emerges from those quantum states of the mind-pixels in the statistical mixtures that have grabbed the attention of the mind through quantum Zeno effect (5) or some non-local means in pre-space-time. Indeed, the many-mind interpretation of quantum theory as proposed by Donald supports this type of formulation (3). Thus, in this scenario, mind does not depend on large quantum coherence to work.

Furthermore, each nuclear spin is a quantum bit, each $O_2$ spin triplet is tightly entangled two-qubit and there exist coherent and incoherent intra-molecular superpositions of multiple nuclear spins under external stimulations (38-40). So, it is plausible that mind could utilize quantum statistical computations similar to those proposed in by Castagnoli and Finkelstein (52). In their model, a triode network made of triplet pairs of spin ½ fermions and their quantum statistical relations due to particle indistinguishability was utilized to implement the computation, together with a randomly-varying magnetic field as heat bath for annealing (52). The scheme develops quantum parallelism through the incoherent superposition of parallel computation paths (i.e., the mixtures). It replaces the superposition of coherent parallel computation paths with the almost indestructible superposition of different permutations of identical particles subject to a given statistics thus surviving decoherence (52).

**Associative memory model**

We have proposed that neural memories are comprised of all possible entangled quantum states of the nuclear spins inside neural membranes and proteins. This proposal calls for extension of the existing associative memory concept in neuroscience to include all possible conformations of neural membranes and proteins in a single neuron (53). A few illustrations are given here. Figure 3 (a) schematically shows a patch of neural membrane containing only the same phospholipids. Such a membrane is much like a blank tape. Figure 3 (b) shows the same neural membrane after cholesterols are added. The changes in membrane configuration are quite noticeable (54-55). These changes can represent memory or information. Figure 3 (c) shows the chemical structure and atomic model of a stearic acid molecule - a saturated fatty acid. Figure 3 (d) shows the chemical structure and atomic model of oleic acid molecule – an unsaturated fatty acid. The only difference between the two fatty acids is that the latter contains a double bond in the middle that causes its kink formation when the double bond is the cis form. When the double bond is in the trans form, the chain is doubly bent so there is no kink. Certainly insert either one of the fatty acids into the membrane shown in Figure 3 (b) would further increase its complexity thus information content. Furthermore, insertions of proteins to neural membranes also significantly change their conformation and dynamics surrounding the inserted proteins (56). Thus, inserting different proteins to neural membranes both in numbers and types can significantly increase the information content of the neural membranes.

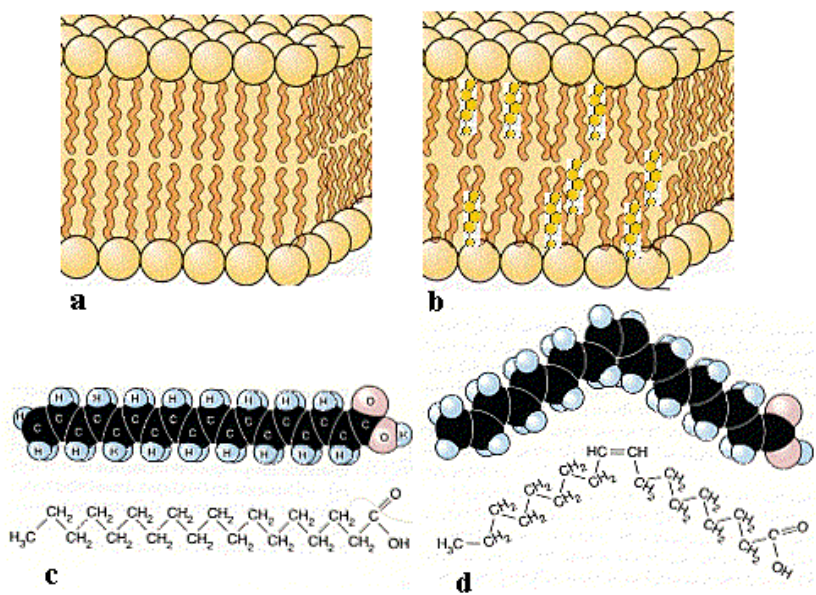

Figure 3. Illustration of neural memory. a shows a neural membrane containing only the same phospholipids. b shows the neural membrane after cholesterols are added. c shows the chemical structure and atomic model of a stearic acid molecule and d shows the chemical structure and atomic model of oleic acid molecule.

### Input and output circuits

We have suggested that action potential modulations of nuclear spin interactions input information to the mind pixels and spin chemistry is the output circuit to classical neural activities. With respect to the input circuit, published studies show that the dynamics of membrane proteins significantly affect the dynamics of surrounding membranes (53,56). Secondly, due to the very small thickness of the neural membrane (~10 nm), even the small voltage (~50 mV) of an action potential can create enormous oscillating electric field inside the neural membranes that in turn can affect the conformations and dynamics of the membrane components such as phospholipid and proteins (57-58). Indeed, the strength and anisotropies of nuclear spin interactions through J-couplings and dipolar couplings are modulated by action potentials as will be shown in a separate paper. Thus, the neural spike trains can directly input information into the mind-pixels made of neural membrane nuclear spins.

Secondly, the weak magnetic field produced collectively by all neural activities may also directly serve as the input. However, the magnitude of said magnetic field is only in the order of $10^{-12}$ Tesla (59). In comparison, $O_2$ and NO can produce a fluctuating local magnetic field as high as a few Tesla according to our own calculations. Thus, the effect of said weak magnetic field on the dynamics of mind-pixels is probably small unless non-linear processes such as stochastic resonance are involved.

Further, we have already pointed out earlier that spin chemistry can serve as the output circuit to classical neural activities because biochemical reactions mediated by free radicals are very sensitive to small changes of magnetic energies as discussed previously. Indeed, many biochemical reactions mediated by radical pairs and biradicals, such as those dual path radical reactions driven/initiated by NO and active oxygen species, have been found to be influenced by the magnetic field in their local environment (19,20). Thus, the functional output of the mind-pixels, being the varying local magnetic field generated by the dynamics of the nuclear spin ensembles as mind-pixels, can directly affect classical neural activities. Further, there may be other mechanisms through which the mind-pixels can influence the classical neural activities of the brain.

**Mechanism of anesthetic action**

As mentioned earlier in this paper, the mechanism of anesthetic action is closely related to the inner workings of consciousness (17). We describe here said mechanism in accordance with our spin-mediated theory. Figure 4 (a) schematically shows the normal diffusion of $O_2$ and NO without anesthetics dissolved into the neural membranes and proteins. As these molecules rapidly diffuse through the membranes, they collide with the neural membrane components and generate strong and fluctuating internal magnetic fields thus activating the nuclear spin ensembles inside these membranes. Figure 4 (b) schematically shows anesthetic perturbations of $O_2$ and NO pathways and neural membranes themselves by anesthetic molecules and the resulting distortion and/or obstruction of these pathways. Such perturbations render $O_2$ and NO not able to perform their normal activation functions thus resulting in unconsciousness.

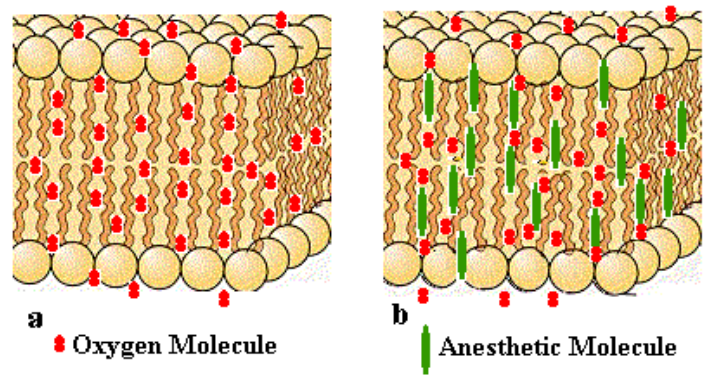

Figure 4. Illustration of anesthetic action. **a** shows the normal diffusion of $O_2$ without anesthetics dissolved into neural membranes. **b** shows anesthetic perturbations of $O_2$ pathways and neural membranes themselves.

## 7. PREDICTIONS AND SUPPORTING EVIDENCE

Several experimentally verifiable predictions can be drawn from our spin-mediated consciousness theory: (1) Significant replacement of $^1H$ with $^2H$ will affect or disrupt consciousness; (2) Significant external disturbances to the dynamics of the nuclear spin ensembles in neural membranes and proteins will interfere with normal conscious functions; (3) Significant drug-induced disturbances to the structure and dynamics of the neural membranes and protein themselves will affect or disrupt consciousness; (4) Significant drug-induced disturbances to the $O_2$ pathways inside the neural membranes will diminish or block consciousness; and (5) Significant lack of $O_2$ in neural membranes will directly affect or disrupt consciousness even if everything else in the brain functions normally. Of course, other predictions and inferences can also be drawn from the present theory. But we will focus our discussions on the above listed a few to see whether there are any experimental evidence supporting these predictions.

With respect to prediction (1), there are published results concerning the biological effects of heavy water ($^2H_2O$ or $D_2O$) that lend indirect support to our proposals (60). Indeed, the only observable physical and chemical difference between regular and heavy water is that the latter has a slightly higher viscosity (60). It was reported that rats restricted to heavy water (99.8%) would drink it freely the first day, then drank progressively less and died within 14 days (62). It was also reported that the incremental increase of the amount of heavy water in regular drinking water lengthened the 24-hour clock of the Circadian rhythm in blind hamsters or normal hamsters kept in constant darkness (62). In principle, these observations can be explained as animals literally loosing their minds because of the loss of their mind-pixels due to gradual $^2H$ substitutions of $^1H$ in neural membranes and proteins. The complication is that the higher viscosity of heavy water might have also contributed to the observed effects. It is possible to design an experimental procedure to either account for or exclude those effects attributable to the higher viscosity.

With respect to prediction (2), there are also quite a few published studies based on transcranial magnetic stimulation ("TMS") that can be at least partially explained based on our theory (63,64), although common wisdom is that TMS induces electrical currents in the brain, causing depolarization of cellular membranes and thereby neural activation (63). It has been found that depending on the locations of stimulation TMS affects the test subject's verbal ability, visualization and other conscious functions (64). According to our theory, TMS can directly affect the dynamics of nuclear spin ensembles in neural membranes and proteins which in turn result in altered, diminished and/or disrupted conscious functions of the brain.

With respect to predictions (3) and (4), many general anesthetics have been found to disturb the structures and dynamics of neural membranes (16,36,37). Thus, the mechanism of their action can be interpreted, according our spin-mediated theory, as caused by their direct effects on $O_2$ pathways and nuclear spin ensembles inside the neural membranes and proteins. Our other paper contains a detailed treatment on anesthetic perturbations of oxygen pathways and membranes themselves (17). Here we will focus on one particularly small anesthetic agent, the nitrous oxide ($N_2O$), also known as the laughing gas. Indeed, the size of $N_2O$ is similar to that of $O_2$ but it does not contain unpaired electrons and is not reactive. It has low polarity that makes it soluble in both water and lipid. Thus, it can be carried to the brain through blood stream and accumulate in the neural membranes. Inhalation of $N_2O$ will cause disorientation, euphoria, numbness and ultimately loss of consciousness if the inhalation dosage is high. The cellular mechanism of these actions by $N_2O$ is so far unknown but seems confined to postsynaptic targets (65). On the other hand, its closely related "cousin" NO contains one unpaired electron and has been discovered as the first small and highly diffusive neural transmitter produced in the brain through enzymatic reactions (66). According to our theory, there indeed exist a natural and straightforward explanation. By dissolving into neural membranes in an inhalation-dose-dependent fashion, $N_2O$ gradually displace $O_2$ in the neural membranes thus diminish or disrupt the activating function of $O_2$.

With respect to prediction (5), it is probably very hard to deprive brain $O_2$ and yet at the same time require its neurons to keep their metabolic functions normal since $O_2$ is an essential component of brain energy production. However, according to our theory in the case of temporary-hypoxia-induced unconsciousness such as that due to sudden loss of air pressure on an airplane, the actually cause may not be the depletion of brain energy resources because of the lack of $O_2$ but the direct loss of $O_2$ as the activating agents.

Now, we briefly turn our attention to the associative memory model proposed herein. There are tens of thousands of research papers on the subject of synaptic plasticity/modification (67). The commonly accepted assumption in neuroscience is that synaptic efficacy is both necessary and sufficient to account for learning and memory (53). Our associative memory model does not conflict with the synaptic efficacy view but extend it to the sub-neural and microscopic domain. Studies show that neural activities modify not only the synaptic efficacy but also the intrinsic properties of the neuron (53).

## 8. DISCUSSIONS AND CONCLUSIONS

Our working philosophy in this paper has been that consciousness is grounded at the bottom of physical reality and emerges from the collective dynamics of known physical candidates inside the brain. We strongly believe that quantum spins are such candidates because they are one of the most fundamental entities in modern physics and, on the other hand, neural membranes and proteins are saturated with spin-carrying nuclei. We have applied reductionist approaches in both physics and neuroscience to reach our tentative conclusions. However, our theory as it stands now is speculative and only qualitatively detailed. We are building a quantitative model and the results will be reported in the near future. We have made important predictions from our theory and presented experimental evidence in support of the same. We have also suggested new experiments to verify the present proposals. Indeed, recent experimental realizations of intra-/inter-molecular nuclear spin coherence and entanglement, macroscopic entanglement of spin ensembles and NMR quantum computation, all in room temperatures, strongly suggest the possibility of a spin-mediated mind.

At this point, one crucial question the reader may ask is how can we explain that cognitive functions seem in general insensitive to environmental and even medical strength external magnetic fields such as those used in MRI? First, the strengths of environmental magnetic fields are in the range of $10^{-4}$ –$10^{-6}$ Tesla (68). In comparison, the internal fluctuating magnetic fields can be as high as several Tesla as we have estimated in a separate paper. Thus, the internal magnetic fields overshadow the environmental ones. But the strengths of magnetic fields used in MRI are in the range of 0.064 to 8.0 Tesla (69) that is comparable to or even higher than the strengths of the internal magnetic fields. So additional explanations are called for. Indeed, the net magnetization of nuclear spins even by magnetic field of several Tesla is only about a few ppm at room temperature (70) which shows that even strong static magnetic fields only have small effects on the thermal dynamics of the neural nuclear spin ensembles. Third, to the extent that said spin ensembles are disturbed by external magnetic fields, it is argued that most of these disturbances do not represent meaningful information to the brain and, further, the brain likely have developed other mechanisms through evolution to counter the effects of most external disturbances. In the cases where external magnetic disturbances were reported to have observable cognitive effects, the above suggestion provides a possible basis for interpreting these effects as the results of said disturbances either too large or containing meaningful information to the brain.

However, we would like to caution that even if our theory is confirmed partly or as a whole by more experiments it will only mark the beginning of a new direction towards a better understanding of consciousness. There are so many questions need to be answered, especially those "hard problems" (11). For example, what are the roles of the nuclear spins carried by different nuclei such as $^1$H, $^{13}$C and $^{31}$P? Do they represent different emotions or feeling of our mind? What are the roles of different biologically available paramagnetic species such as $O_2$ and neural transmitter NO? Do they activate different perceptions of mind?

In conclusion, we have proposed a novel theory of consciousness in which quantum spins play the central role as mind-pixels and the unity of mind is achieved by entanglement of these mind-pixels. To justify such a choice, we have shown that spin is the origin of quantum effects in both Bohm and Hestenes quantum formulism and a fundamental quantum process associated with the structure of space-time. Applying these ideas to the particular structures and dynamics of the brain, we have theorized how consciousness emerges from the collapse of the decoherence-resistant entangled spin states via contextual, non-computable and irreversible processes. We have suggested that these entangled spin states are formed through action potential modulated nuclear spin interactions and paramagnetic $O_2$/NO driven activations and survive rapid decoherence through quantum Zeno effects or in decoherence-free subspaces We have further suggested that the collective spin dynamics associated with said collapses have effects through spin chemistry on classical neural activities thus influencing the neural networks of the brain. Our proposals imply the extension of associative encoding of neural memories to the dynamical structures of neural membranes and proteins. Therefore, according our theory the neural substrates of consciousness are comprised of the following: (a) nuclear spin ensembles embedded in neural membranes and proteins which serve as the "mind-screen" with nuclear spins as the pixels, (b) the neural membranes and proteins themselves which serve as the matrices for the mind-screen and neural memories; and (c) the biologically available paramagnetic species such as $O_2$ and NO which serve as the pixel-activating agents.

Part II.    **Photon Induced Non-local Effects of General Anesthetics on the Brain**

## 1. INTRODUCTION

Photons are intrinsically quantum objects and natural long-distance carriers of information in both classical and quantum communications (1). Since brain functions involve information and many experiments have shown that quantum entanglement is physically real, we have contemplated from the perspective of our herein described hypothesis (2) on the possibility of entangling the quantum entities inside the brain with those in an external anesthetic sample and carried out experiments toward that end. Indeed, Quantum entanglement is ubiquitous in the microscopic world and manifests itself macroscopically under some circumstances (3, 4). Further, quantum spins of electrons and photons have now been successfully entangled in various ways for the purposes of quantum computation, memory and communication (5, 6). In the field of neuroscience, we have suggested that nuclear and/or electronic spins inside the brain may play important roles in certain aspects of brain functions such as perception (2). Arguably, we could test our hypothesis by first attempting to entangle these spins with those of a chemical substance such as a general anesthetic and then observing the resulting brain effects such attempt may produce, if any. Indeed, instead of armchair debate on how the suggested experiments might not work, we just went ahead and carried out the experiments over the periods of more than a year. Here, we report our results. We point out from the outset that although it is commonly believed that quantum entanglement alone cannot be used to transmit classical information, the function of the brain may not be totally based on classical information (2).

Here we report that applying magnetic pulses to the brain when a general anesthetic sample was placed in between caused the brain to feel the effect of said anesthetic for several hours after the treatment as if the test subject had actually inhaled the same. The said effect is consistently reproducible on all four subjects tested. We further found that drinking water exposed to magnetic pulses, laser light, microwave or even flashlight when an anesthetic sample was placed in between also causes consistently reproducible brain effects in various degrees. We have in addition tested several medications including morphine and obtained consistently reproducible results. Further, through additional experiments we have verified that the said brain effect is the consequence of quantum entanglement between quantum entities inside the brain and those of the chemical substance under study induced by the photons of the magnetic pulses or applied lights. We suggest that the said quantum entities inside the brain are nuclear and/or electron spins and discuss the profound implications of these results.

## 2. METHODS, TEST SUBJECTS & MATERIALS

Figure 1A shows a typical setup for the first set of experiments. It includes a magnetic coil with an estimated 20*W* output placed at one inch above the right side of a test subject's forehead, a small flat glass-container inserted between the magnetic coil and the forehead, and an audio system with adjustable power output and frequency spectrum controls connected to the magnetic coil. When music is played on the audio system, the said magnetic coil produces magnetic pulses with frequencies in the range of 5Hz to10kHz. Experiments were conducted with said container being filled with different general anaesthetics, medications, or nothing/water as control, and the test subject being exposed to the magnetic pulses for 10*min* and not being told the content in the container or details of the experiments. The indicators used to measure the brain effect of said treatment were the first-person experiences of any unusual sensations such as numbness, drowsiness and/or euphoria which the subject felt after the treatment and the relative degrees of these unusual sensations on a scale of 10 with 0=nothing, 1=weak, 2=light moderate, 3=moderate, 4=light strong, 5=strong, 6=heavily strong, 7=very strong, 8=intensely strong, 9=extremely strong and 10=intolerable. The durations of the unusual sensations and other symptoms after the treatment such as nausea or headache were also recorded.

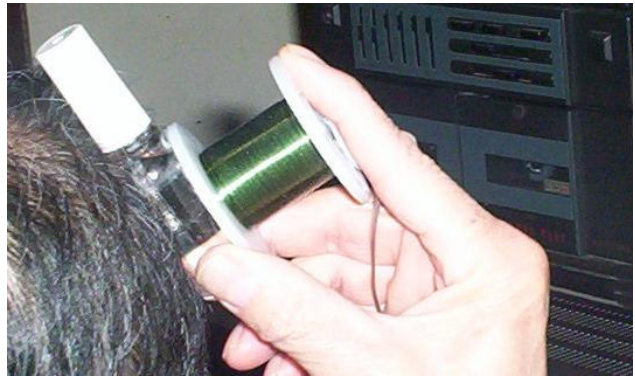

Figure 1A. Photograph of a typical setup for the first set of experiments with a magnetic coil.

Figure 2A shows a typical setup for the second set of experiments. It includes the magnetic coil connected to the audio system, a large flat glass-container filled with 200*ml* fresh tap water and the small flat glass-container inserted between the magnetic coil and larger glass-container. Figure 3A shows a typical setup for the second set of experiments when a red laser with a 50*mW* output and wavelengths of 635*nm* – 675*nm* was used. All Experiments were conducted in the dark with the small flat glass-container being filled with different general anaesthetics, medications, or nothing/water as control, the large glass-container being filled with 200*ml* fresh tap water and exposed to the magnetic pulses or laser light for 30*min* and the test subject consuming the treated tap water but not being told the content in the small container or details of the experiments. The indicators used for measuring the brain effects were the same as those used in the first set of experiments. Experiments were also carried out respectively with a 1200*W* microwave oven and a flashlight powered by two size-D batteries. When the microwave oven was used, a glass tube containing 20*ml* fresh tap water was submerged into a larger glass tube containing 50*ml* general anaesthetic and exposed to microwave radiation for 5*sec*. The said procedure was repeated for multiple times to collect a total of 200*ml* treated tap water for consumption. When the flashlight was used, the magnetic coil shown in Figure 2 was replaced with the flashlight.

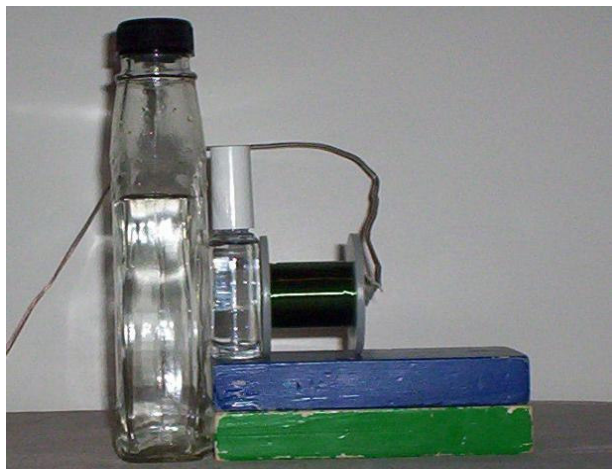

Figure 2A. Photograph of a typical setup for the second set of experiments with a magnetic coil.

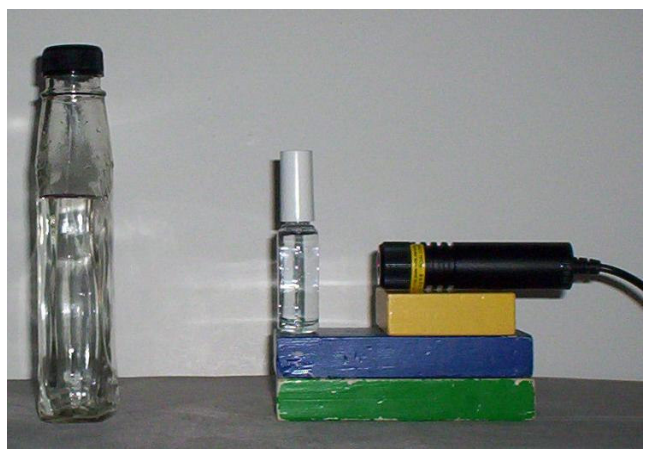

Figure 3A. Photograph of a typical setup for the second set of experiments with a 50*mW* red laser device.

To verify that the brain effects experienced by the test subjects were the consequences of quantum entanglement between quantum entities inside the brain and those in the chemical substances under study, the following additional experiments were carried out. Figure 4A shows a typical setup of the entanglement verification experiments. The setup is the reverse of the setup shown in Figure 3A. In addition, the small flat glass-container with a chemical substance or nothing/water as control was positioned with an angle to the incoming laser light to prevent reflected laser light from re-entering the large glass-container.

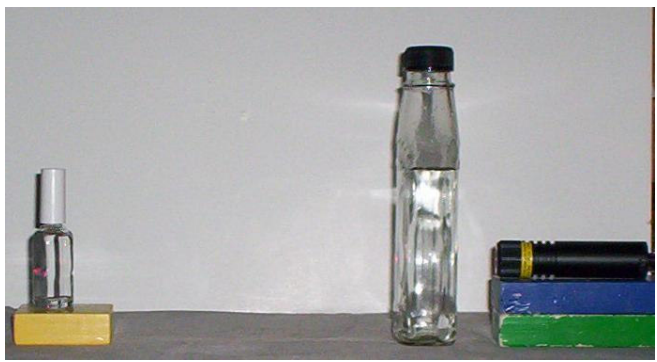

Figure 4A. Photograph of a setup for the entanglement verification experiments.

In the first set of entanglement verification experiments, the laser light from the red laser first passed through the large glass-container with 200*ml* fresh tap water and then through the small flat glass-container filled with a chemical substance or nothing/water as control located about 300*cm* away. After 30*min* of exposure to the laser light, a test subject consumed the exposed tap water without being told the content in the small container or details of the experiments and reported the brain effects felt for the next several hours.

In the second set of entanglement verification experiments, 400*ml* fresh tap water in a glass-container was first exposed to the radiation of the magnetic coil for 30*min* or that of the 1500*W* microwave oven for 2*min*. Then the test subject immediately consumed one-half of the water so exposed. After 30*min* from the time of consumption the other half was exposed to magnetic pulses or laser light for 30 minutes using the setup shown in Figure 2 and Figure 4 respectively. The test subject reported, without being told the content in the small container or details of the experiments, the brain effects felt for the whole period from the time of consumption to several hours after the exposure had stopped.

In the third set of entanglement verification experiments, one-half of 400*ml* Poland Spring water with a shelve time of at least three months was immediately consumed by the test subject. After 30*min* from the time of consumption the other half was exposed to the magnetic pulses or laser light for 30*min* using the setup shown in Figure 2A and Figure 4A respectively. Test subject reported, without being told the content in the small container or details of the experiments, the brain effects felt for the whole period from the time of consumption to several hours after the exposure had stopped.

In the fourth set of entanglement verification experiments, the test subject would take one-half of the 400*ml* fresh tap water exposed to microwave for 2*min* or magnetic pulses for 30*min* to his/her workplace located more than 50 miles away (in one case to Beijing located more than 6,500 miles away) and consumed the same at the workplace at a specified time. After 30*min* from the time of consumption, the other half was exposed to magnetic pulses or laser light for 30*min* at the original location using the setup shown in Figure 2A and Figure 4A respectively. The test subject reported the brain effects felt without being told the content in the small container or details of the experiments for the whole period from the time of consumption to several hours after the exposure had stopped.

With respect to the test subjects, Subject A and C are respectively the first author and co-author of this paper and Subject B and C are respectively the father and mother of the first author. All four test subjects voluntarily consented to the proposed experiments. To ensure safety, all initial experiments were conducted on Subject A by himself. Further, all general anaesthetics used in the study were properly obtained for research purposes and all medications were either leftover items originally prescribed to Subject C's late mother or items available over the counter. To achieve proper control, repeating experiments on Subject A were carried out by either Subject B or C in blind settings, that is, he was not told whether or what general anaesthetic or medication were applied before the end of the experiments. Further, all experiments on Subject B, C and D were also carried out in blind settings, that is, these test subjects were not told about the details of the experiments on them or whether or what general anaesthetic or medication were applied.

## 3. RESULTS

Table 1 summarizes the results obtained from the first two sets of experiments described above and Table 2 breakdowns the summary into each general anaesthetic studied plus morphine in the case of medications. In the control studies for the first set of experiments, all test subjects did not feel anything unusual from the exposure to magnetic pulses except vague or weak local sensation near the site of exposure. In contrast, all general anaesthetics studied produced clear and completely reproducible brain effects in various degrees and durations as if the test subjects had actually inhaled the same. These brain effects were first localized near the site of treatment and then spread over the whole brain and faded away within several hours. But residual brain effects (hangover) lingered on for more than 12 hours in most cases. Among the general anaesthetics studied, chloroform and deuterated chloroform (chloroform D) produced the most pronounced and potent brain effects in strength and duration followed by isoflorance and diethyl ether. Tribromoethanol dissolved in water (1:50 by weight) and ethanol also produced noticeable effects but they are not summarized in the table.

As also shown in Table 1, while the test subjects did not feel anything unusual from consuming the tab water treated in the control experiments with magnetic pulses or laser light, all the general anaesthetics studied produced clear and completely reproducible brain effects in various degrees and durations respectively similar to the observations in the first set of experiments. These effects were over the whole brain, intensified within the first half hour after the test subjects consumed the treated water and then faded away within the next a few hours. But residual brain effects lingered on for more than 12 hours as in the first set of experiments. Among the general anaesthetics studied, again chloroform and deuterated chloroform produced the most pronounced and potent effect in strength and duration followed by isoflorance and diethyl ether as illustrated in Figure 5A. Tribromoethanol dissolved in water (1:50 by weight) and ethanol also produced noticeable effects but they are not summarized in the table.

**Table1. Summary of results obtained from the first two sets of experiments**

| Table 1 | 1st Set: Magn. Coil | | 2nd Set: Magn. Coil | | Laser Light | | Flashlight | | Microwave | |
|---|---|---|---|---|---|---|---|---|---|---|
| | Test # | Effect | Test # | Effect | Test # | Effect | Test # | Effect | Test # | Effect |
| Anaesthetics | | | | | | | | | | |
| Subject A | 13 | Yes | 16 | Yes | 22 | Yes | 8 | Yes | 3 | Yes |
| Subject B | 2 | Yes | 2 | Yes | 3 | Yes | 0 | N/A | 1 | Yes |
| Subject C | 2 | Yes | 6 | Yes | 6 | Yes | 0 | N/A | 1 | Yes |
| Subject D | 2 | Yes | 1 | Yes | 5 | Yes | 0 | N/A | 0 | N/A |
| Medications | | | | | | | | | | |
| Subject A | 17 | Yes | 14 | Yes | 16 | Yes | 1 | Yes | 3 | Yes |
| Subject B | 1 | Yes | 1 | Yes | 3 | Yes | 0 | N/A | 2 | Yes |
| Subject C | 3 | Yes | 1 | Yes | 4 | Yes | 0 | N/A | 1 | Yes |
| Subject D | 0 | N/A | 0 | N/A | 3 | Yes | 0 | N/A | 1 | Yes |
| Control | | | | | | | | | | |
| Subject A | 12 | No | 5 | No | 11 | No | | | | |
| Subject B | 3 | No | 0 | N/A | 1 | No | | | | |
| Subject C | 1 | No | 2 | No | 4 | No | | | | |
| Subject D | 0 | N/A | 0 | N/A | 1 | No | | | | |

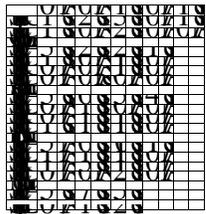

**Table2. Breakdowns of the summary in Table 1 into each general anesthetic studied plus morphine in the case of medications.**

| Table 2 | 1st Set: Magn. Coil | | 2nd Set: Mag. Coil | | Red Laser | | Flashlight | | Microwave | |
|---|---|---|---|---|---|---|---|---|---|---|
| | Test # | Effect | Test # | Effect | Test # | Effect | Test # | Effect | Effect | Effect |
| Chloroform | | | | | | | | | | |
| Subject A | 2 | Yes | 2 | Yes | 5 | Yes | 2 | Yes | 3 | Yes |

| Subject C | 0 | N/A | 1 | Yes | 2 | Yes | | | | |
|---|---|---|---|---|---|---|---|---|---|---|
| Subject D | 0 | N/A | 0 | N/A | 2 | Yes | | | | |
| Other Med | | | | | | | | | | |
| Subject A | 7 | Yes | 4 | Yes | | | | | | |
| Subject B | 1 | Yes | 0 | N/A | | | | | | |
| Subject C | 3 | Yes | 0 | N/A | | | | | | |
| Subject D | 0 | N/A | 0 | N/A | | | | | | |

In addition, available results with flashlight and microwave as photon sources are also summarized in Table 1 respectively. In both cases general anaesthetics studied produced clear and reproducible brain effects. But the brain effects produced with microwave exposure were much stronger than those by flashlight.

Table 1 also summarizes results obtained with several medications including morphine, fentanyl, oxycodone, nicotine and caffeine in first and second sets of experiments. We found that they all produced clear and completely reproducible brain effects such as euphoria and/or hastened alertness in various degrees and durations respectively. For example, in the case of morphine in the first set of experiments the brain effect was first localized near the site of treatment and then spread over the whole brain and faded away within several hours. In the case of morphine in the second set of experiments the brain effect was over the whole brain, first intensified within the first half hour after the test subjects consumed the treated water and then faded away within the next a few hours as illustrated in Figure 6A.

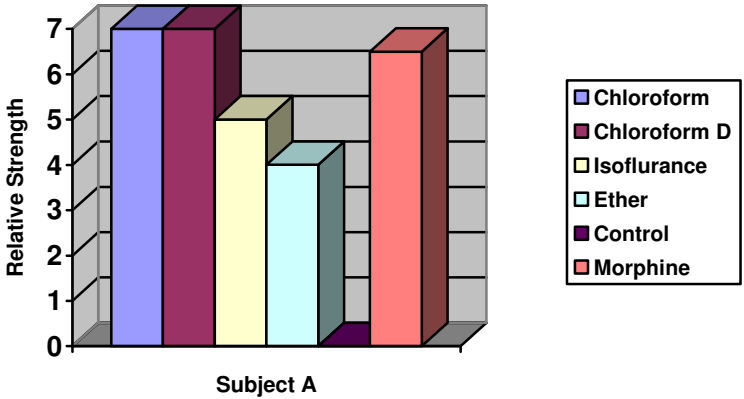

**Figure 5A. Brain Effects of General Anaesthetics and Morphine**

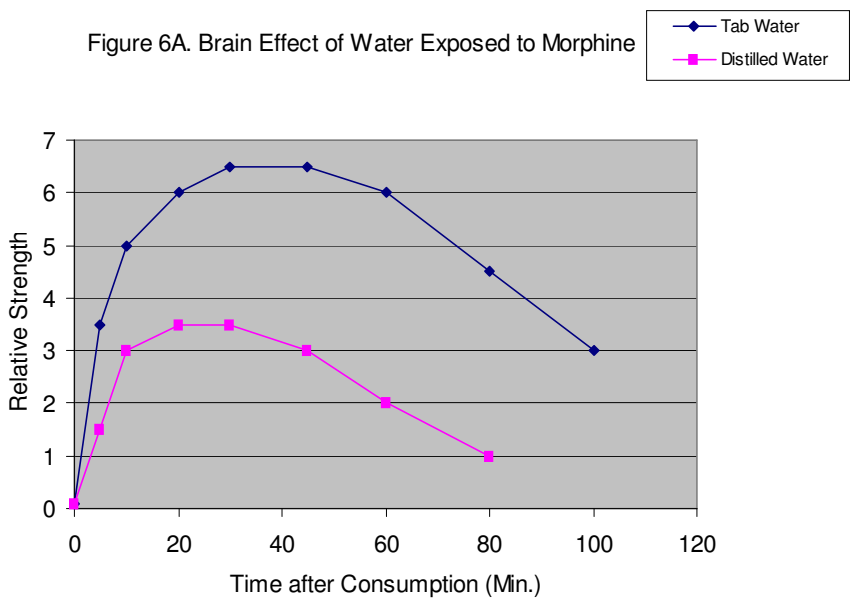
Figure 6A. Brain Effect of Water Exposed to Morphine

Comparative experiments were also conducted on Subject A and C with chloroform and diethyl ether by asking them to inhale the vapours of chloroform and diethyl respectively for 5*sec* and compare the brain effect felt with those in the two sets of experiments described above. The brain effects induced in these comparative experiments were qualitatively similar to those produced in various experiments described above when chloroform and diethyl ether were respectively used for the exposure to photons of various sources.

Furthermore, through additional experiments we also made the following preliminary observations. First, the brain effects in the first set of experiments could not be induced by a permanent magnet in the place of the magnetic coil. Nor could these effects be produced by a third magnetic coil placed directly above the head of the test subject and connected to a second magnetic coil through an amplifier with the second magnetic coil receiving magnetic pulses from a first magnetic coil after the said magnetic pulses first passed through the anaesthetic sample. That is, the brain effects could not be transmitted through an electric wire. Second, in the second set of experiments the water exposed to magnetic pulses, laser light, microwave and flashlight when a chemical substance was present tasted about the same as that before the exposure. Third, heating tap water exposed to magnetic pulses or laser light in the presence of a chemical substance diminished the brain effect of the said substance. Fourth, when distilled water was used instead of fresh tap water the observed brain effects were markedly reduced as illustrated in Figure 6 in the case of morphine.

Table 3 summarizes the results obtained with the entanglement verification experiments carried out so far with chloroform, deuterated chloroform, diethyl ether and morphine. With all four sets of experiments, clear and consistently reproducible brain effects were experienced by the test subjects above and beyond what were noticeable in the control portions of the experiments under blind settings.

**Table3. Summary of the results obtained with the entanglement verification experiments carried out so far with chloroform, deuterated chloroform, diethyl ether and morphine.**

| Table 3 | First Set | | Second Set | | Third Set | | Fourth Set | |
|---|---|---|---|---|---|---|---|---|
| | Test # | Effect | Test # | Effect | Test # | Effect | Test # | Effect |
| Subject A | 8 | Yes | 8 | Yes | 3 | Yes | 3 | Yes |
| Subject B | 2 | Yes | 3 | Yes | 2 | Yes | 1 | Yes |
| Subject C | 3 | Yes | 2 | Yes | 1 | Yes | 1 | Yes |
| Control | | | | | | | | |
| Subject A | 2 | No | 8 | No | 3 | No | 3 | No |
| Subject B | 0 | N/A | 3 | No | 2 | No | 1 | No |
| Subject C | 1 | No | 2 | No | 1 | No | 1 | No |

More specifically, in the first set of entanglement verification experiments, the brain effects experienced by the test subjects were the same as those in which the setup shown in Figure 3A was used. In the second, third and fourth sets of these experiments, all test subjects did not feel anything unusual in the first half hour after consuming the first half of the water either exposed to microwave/magnetic pulses or just sit on the shelf for more than 3 months. But within minutes after the second half of the same water was exposed to the laser light or magnetic pulses in the presence of general anaesthetics, the test subjects would experience clear and completely reproducible brain effect of various intensities as if they have actually inhaled the general anaesthetic used in the exposure of the second half of the water. The said brain effects were over the whole brain, first intensified within minutes after the exposure began and persisted for the duration of the said exposure and for the next several hours after the exposure had stopped. Further, all other conditions being the same, magnetic coil produced more intense brain effects than the red laser. Furthermore, all other conditions being the same, the water exposed to microwave or magnetic pulses before consumption produced more intense brain effects than water just sitting on the shelf for more than 3 months before consumption.

## 3. DISCUSSIONS & CONCLUSIONS

With respect to the second, third and fourth sets of entanglement verification experiments, the only possible explanation for the brain effects experienced by the test subjects are that they were the consequences of quantum entanglement because the water consumed by the test subjects was never directly exposed to the magnetic pulses or the laser lights in the presence of the chemical substances.

There are other indications that quantum entanglement was the cause of the brain effects experienced by the test subjects. First, the brain effect inducing mean could not be transmitted through an electrical wire as already reported above. Second, the said inducing mean did not depend on the wavelengths of the photons generated. Thus, mere interactions among the photons, a chemical substance and water will induce brain effects after a test subject consumes the water so interacted. While designing and conducting the herein described experiments, the first author became aware of the claims related to the so called "water memory" (7). However, since these claims were said to be non-reproducible, we do not wish to discuss them further here except to say that we currently do not subscribe to any of the existing views on the subject and readers are encouraged to read our recent online paper on quantum entanglement (8).

In light of the results from the entanglement verification experiments, we conclude that the brain effects experienced by the test subjects were the consequences of quantum entanglement between quantum entities inside the brains and those of the chemical substances under study induced by the entangling photons of the magnetic pulses or applied lights. More specifically, the results obtained in the first set of experiments can be interpreted as the consequence of quantum entanglement between the quantum entities in the brain and those in the chemical substances induced by the photons of the magnetic pulses. Similarly, the results obtained from the second sets of experiments can be explained as quantum entanglement between the quantum entities in the chemical substance and those in the water induced by the photons of the magnetic pulses, laser light, microwave or flashlight and the subsequent physical transport of the water entangled with the said chemical substance to the brain after consumption by the test subject which, in turn, produced the observed brain effects through the entanglement of the quantum entities inside the brain with those in the consumed water.

We would like to point out that although the indicators used to measure the brain effects were qualitative and subjective, they reflect the first-person experiences of the qualities, intensities and durations of these effects by the test subjects since their brains were directly used as experimental probes. Further, these effects are completely reproducible under blind experimental settings so that possible placebo effects were excluded. However, as with many other important new results, replications by others are the key to independently confirm our results reported here. Our experiments may appear simple and even "primitive" but the results and implications are profound.

We first chose general anaesthetics in our experiments because they are among the most powerful brain-influencing substances. Our expectation was that, if nuclear and/or electronic spins inside the brain are involved in brain functions such as perception as recently hypothesized by us[2], the brain may be able to sense the effect of an external anaesthetic sample through quantum entanglement between these spins inside the brain and those of the said anaesthetic sample induced by the photons of the magnetic pulses by first interacting with the nuclear and/or electronic spins inside the said anaesthetic sample, thus carrying quantum information about the anaesthetic molecules, and then interacting with the nuclear and/or electronic spins inside the brain.

We suggest here that the said quantum entities inside the brains are likely nuclear and/or electronic spins for the reasons discussed below. Neural membranes and proteins contain vast numbers of nuclear spins such as $^1$H, $^{13}$C, $^{31}$P and $^{15}$N. These nuclear spins and unpaired electronic spins are the natural targets of interaction with the photons of the magnetic pulses or other sources. These spins form complex intra- and inter-molecular networks through various intra-molecular J- and dipolar couplings and both short- and long-range intermolecular dipolar couplings. Further, nuclear spins have relatively long relaxation times after excitations (9). Thus, when a nematic liquid crystal is irradiated with multi-frequency pulse magnetic fields, its $^1$H spins can form long-lived intra-molecular quantum coherence with entanglement for information storage (10). Long-lived (~ .05 *ms*) entanglement of two macroscopic electron spin ensembles in room temperature has also been achieved (3). Furthermore, spin is a fundamental quantum process with intrinsic connection to the structure of space-time (11) and was shown to be responsible for the quantum effects in both Hestenes and Bohmian quantum mechanics (12, 13). Thus, we have recently suggested that these spins could be involved in brain functions at a more fundamental level (2).

Several important conclusions and implications can be drawn from our findings. First, biologically/chemically meaningful information can be transmitted through quantum entanglement from one place to another by photons and possibly other quantum objects such as electrons, atoms and even molecules. Second, both classical and quantum information can be transmitted between locations of arbitrary distances through quantum entanglement alone. Third, instantaneous signalling is physically real which implies that Einstein's theory of relativity is in real (not just superficial) conflict with quantum theory. Fourth, brain processes such as perception and other biological processes likely involve quantum information and nuclear and/or electronic spins may play important roles in these processes. Further, our findings provide important new insights into the essence and implications of the mysterious quantum entanglement and clues for solving the long-standing measurement problem in quantum theory including the roles of the observer and/or consciousness. Very importantly, our findings also provide a unified scientific framework for explaining many paranormal and/or anomalous effects such as telepathy, telekinesis and homeopathy, if they do indeed exist, thus transforming these paranormal and/or anomalous effects into the domains of conventional sciences. With respect applications, our findings enable various quantum entanglement technologies be developed. Some of these technologies can be used to deliver the therapeutic effects of many drugs to various biological systems such as human bodies without physically administrating the same to the said systems. This will dramatically reduce waste and increase productivity because the same drugs can be repeatedly used to deliver their therapeutic effects to the mass on site or from remote locations of arbitrary distances. Further, many substances of nutritional and recreational values can be repeatedly administrated to desired biological systems such as human bodies through the said technologies either on site or from remote locations. Other such technologies can be used for instantaneous communications between remote locations of arbitrary distances in various ways. Potentially, these technologies can also be used to entangle two or more human minds for legitimate and beneficial purposes.

In addition, several important conclusions and implications can be drawn from our experimental findings. First, biologically/chemically meaningful information can be transmitted through quantum entanglement from one place to another by photons and possibly other quantum objects such as electrons, atoms and even molecules. Second, both classical and quantum information can be transmitted between locations of arbitrary distances through quantum entanglement alone. Third, instantaneous signaling is physically real which implies that Einstein's theory of relativity is in real (not just superficial) conflict with quantum theory. Fourth, brain processes such as perception and other biological processes likely involve quantum information and nuclear and/or electronic spins may play important roles in these processes. Further, our findings provide important new insights into the essence and implications of the mysterious quantum entanglement and clues for solving the long-standing measurement problem in quantum theory including the roles of the observer and/or consciousness. Very importantly, our findings also provide a unified scientific framework for explaining many paranormal and/or anomalous effects such as telepathy, telekinesis and homeopathy, if they do indeed exist, thus transforming these paranormal and/or anomalous effects into the domains of conventional sciences.

Finally, with respect applications, our experimental findings enable various quantum entanglement technologies be developed. Some of these technologies can be used to deliver the therapeutic effects of many drugs to various biological systems such as human bodies without physically administrating the same to the said systems. This will dramatically reduce waste and increase productivity because the same drugs can be repeatedly used to deliver their therapeutic effects to the mass on site or from remote locations of arbitrary distances. Further, many substances of nutritional and recreational values can be repeatedly administrated to desired biological systems such as human bodies through the said technologies either on site or from remote locations. Other such technologies can be used for instantaneous communications between remote locations of arbitrary distances in various ways. Potentially, these technologies can also be used to entangle two or more human minds for legitimate and beneficial purposes.

## Part III. Evidence of Non-local Chemical, Thermal and Gravitational Effects

### 1. INTRODUCTION

Scientific methods require that one conform one's knowledge of nature to repeatable observations. Thus, it is unscientific to reject what's observed repeatedly and consistently. With this in mind, we comment that quantum entanglement has been recently shown to be physically real in many laboratories (1, 2). Indeed, quantum entanglement is ubiquitous in the microscopic world and manifests itself macroscopically under some circumstances (1, 2). But common belief is that it alone cannot be used to transmit information3 nor could it be used to produce macroscopic non-local effects. Yet we have recently found evidence of non-local effects of chemical substances on the brain produced through it (4, 5). Indeed, spins of electrons, photons and nuclei have now been successfully entangled in various ways for the purposes of quantum computation and communication (6, 7). While our reported results are under independent verifications by other groups, we report here our experimental findings of non-local chemical, thermal and gravitational effects in simple physical systems such as reservoirs of water quantum-entangled with water being manipulated in a remote reservoir (4). With the aids of high-precision instruments, we have found that the pH value, temperature and gravity of water in the detecting reservoirs can be non-locally affected through manipulating water in the remote reservoir. In particular, the pH value changes in the same direction as that being manipulated; the temperature can change against that of local environment; and the gravity can change against local gravity. These non-local effects are all reproducible and can be used for non-local signaling and many other purposes. We suggest that they are mediated by quantum entanglement between nuclear and/or electron spins in treated water and discuss the profound implications of these results.

Our motivation for measuring pH change of water in one reservoir, while manipulating water in a remote reservoir quantum-entangled with the former, is to investigate whether and how pH value in the water being measured shifts under non-local influences. Our motivation for measuring temperature variation of water in one reservoir, while manipulating water in a remote reservoir quantum-entangled with the former, is to investigate whether and how the thermodynamics of water being measured changes under non-local influences. Our motivation for measuring gravity change of one reservoir of water, while manipulating water in a remote reservoir quantum-entangled with the former, is to investigate whether gravity is a non-local effect associated with quantum entanglement.

The successes of the experiments described herein were achieved with the aids of high-precision analytical instruments. They include an Ohaus Voyager Analytical Balance with capacity 210g, resolution 0.1mg, repeatability 0.1mg and sensitivity drift 3 PPM/ºC, a Control Company traceable-calibration digital thermometer with resolution 0.001ºC and repeatability 0.002ºC near 25ºC in liquid such as water (estimated from calibration data provided), and a Hanna microprocessor pH meter Model 213 with resolution 0.001 and repeatability 0.002. The other key apparatus is a 25-litre Dewar filled with liquid nitrogen and positioned remotely at a desired distance which not only provided the drastic changes in the water being manipulated but also served as a natural Faraday cage blocking any possible electromagnetic influence between the water being measured and the water being manipulated. Also vital to the success of the experiments described herein was the stable environment found in an underground room which shields many external noises such as mechanical vibration, air turbulence and large temperature change.

## 2. MATERIALS & METHODS

Quantum-entangled stock water in individual volumes of 500ml or similar quantities was prepared as described previously (4) which might then be split into smaller volumes or combined into larger ones based on needs. Briefly, in one procedure 500ml fresh tap water in a closed plastic reservoir was exposed to microwave radiation in a 1500W microwave oven for 2min and then left in room temperature for 24 hours before use. In a second procedure 500ml fresh tap water in the closed plastic reservoir was exposed to audio-frequency radiations of a 20W magnetic coil for 30min and then left in room temperature for 24 hours before use. In a third procedure, 500ml bottled natural water was simply left in room temperature for at least 30 days before use. In a fourth procedure, 500ml bottled distilled water was simply left in room temperature for at least 30 days before use. It was found previously that the stock water prepared according to these procedures is quantum-entangled (4).

Figure 1 shows a diagram of the key experimental setup. It includes (a) the analytical balance calibrated internally and stabilized in the underground room for more than one week before use and a tightly closed plastic first-reservoir containing 175ml water split from the 500ml stock water which is placed on the wind-shielded pan of the balance with 1-inch white foam in between as insulation; (b) the digital thermometer and calibrated pH meter placed into the middle of a glass second-reservoir containing 75ml water split from the 500ml stock water which is closed to prevent air exchange; and (c) the 25-litre Dewar containing 15-25 litres of liquid nitrogen which is located at a distant of 50 feet from the underground room and a tightly closed plastic third-reservoir containing 250ml water split from the 500ml stock water to be submerged into the liquid nitrogen in the Dewar at a specified time.

Experiments with the above first-setup were carried out as follows: (a) prepare the 500ml quantum-entangled stock water, divide the same into 175ml, 75ml and 250ml portions and put them into their respective reservoirs described above; (b) set up the experiment according to Figure 1 and let the instruments to stabilize for 30min before any measurements is taken; (c) record for 20min minute-by-minute changes of pH value and temperature of the water in the first-reservoir and weight of the second-reservoir with water before submerging the third reservoir into liquid nitrogen; (d) submerge the third-reservoir with water into liquid nitrogen for 15min or another desired length of time and record the instrument readings as before; and (e) take the third-reservoir out of liquid nitrogen, thaw the same in warm water for 30min or longer and, at the same time, record the instrument readings as before. Control experiments were carried out in same steps with nothing done to the water in the third-reservoir.

In one variation of the above setup, the closed plastic third-reservoir was replaced with a metal container and instead of freeze-thaw treatment the water in the metal container was quickly heated to boiling within 4-5 minutes and then cooled in cold water. In a second variation of the above setup, the gravity portion of the experiment was eliminated and the water in the first and second reservoirs was combined into a closed thermal flask which prevents heat exchange between the water being measured and its local environment. In a third variation of the above setup, the gravity portion of the experiment was eliminated and the water in the first and second reservoirs was combined into a fourth plastic container in which 5ml concentrated HCl (30% by weight) was first added, then 20g NaOH powder was added and next the same water was transferred to a metal container and heated to boiling on a stove. In a fourth variation of the above first-setup, the 25-litre Dewar containing liquid nitrogen was replaced by a large water tank located 20-feet above the underground room which contained 200-gallon tap water sitting in room temperature for months and, instead of submersion, the water in the third-reservoir was poured into the large water tank the purpose of which was to quantum-entangle the poured water with the water in the large tank. In a fifth variation of the above setup, the gravity portion of the experiment was eliminated and the water in the first and second reservoirs was combined into a closed glass fourth-reservoir which was moved to a location more than 50 miles away from the Dewar for temperature measurement.

Figure 2 shows a diagram of the second experimental setup. It includes: (a) a red laser with a 50mW output and wavelengths 635nm – 675nm placed next and pointed to a flat glass first-reservoir containing 200ml tap water sitting in room temperature for more than a week without air exchange; (b) the calibrated pH meter and optionally the digital thermometer placed into the middle of the said flat glass reservoir which was closed to prevent air exchange; and (3) a round glass second-reservoir containing 100ml concentrated HCl (30% by weight) to be placed 500cm away from the first-reservoir at a specified time.

Experiments with the above second setup were carried out as follows: (a) prepare the 200ml tap water and set up the experiment according Figure 2; turn on the laser so that the laser light first passes through the first-reservoir and then gets scattered on a nearby concrete wall, and let the instruments to stabilize for 30min before any measurement is taken; (b) record for 10min. minute-by-minute changes of pH value and optionally temperature of the water in the first-reservoir; and (c) place the second reservoir containing 100ml HCl on the path of the laser light and at a distance of 500cm from the first reservoir and record for 60min or longer instrument readings as before. Control experiments were carried out in same steps in the absence of HCl.

## 3. RESULTS

Figures 3, 4 and 5 summarize the results obtained from experiments conducted with the key setup and one batch of quantum-entangled water which were simply bottled natural water with a shelf time of more than 90 days. Similar results were also obtained with water prepared according to other quantum entanglement methods mentioned above and other quantum-entangled liquid such as olive oil, alcohol and even Coke Cola as discussed later. The different distances of the Dewar from the underground room where most measurements were done made no noticeable differences with respect to the results obtained.

Figure 3 shows changes of pH value of the water in the second-reservoir during the three stages of manipulations of the water in the remote third-reservoir. As shown, within minutes after the remote third-reservoir was submerged into liquid nitrogen, during which the temperature of water being manipulated would drop from about 25ºC to -193 ºC, the pH value of the water in the second reservoir steadily stopped dropping and then started rising, but about 20 minutes after the frozen water was taken out of liquid nitrogen and thawed in warm water the pH value of the same steadily leveled off and started dropping again. In contrast, the control experiments did not show such dynamics. It is known that the pH value of water increases as its temperature goes down to 0ºC. Therefore, the pH value of water being measured goes in the same direction as the remote water when the latter is manipulated. The difference in pH values from control in which no freeze-thaw was done at the point of thawing is about 0.010. However, if the water being measured is kept in a thermal flask to prevent heat exchange with the local environment, no effect on pH value was observed under freeze-thaw treatment of the remote water. Statistical analysis performed on data collected after freezing for 10 minutes show that the results are significantly different under these different treatments/settings.

Figure 4 shows temperature variations of the water in the second-reservoir during the three stages of manipulations of the water in the remote third-reservoir. As shown, before the submersion of the remote third-reservoir into liquid nitrogen the temperature of the water in the second-reservoir rose in small increments due to, by design, the slight temperature difference between the local environment and the water inside the second reservoir; but within about 4-5 minutes after the remote third-reservoir was submerged into liquid nitrogen, during which the temperature of water being manipulated would drop from about 25ºC to -193 ºC, the temperature of the water in the second reservoir first stopped rising and then steadily dropped in small increments; and then within about 4-5 minutes after the frozen water was taken out of liquid nitrogen and thawed in warm water the temperature of the same first stopped dropping and then steadily rose again in small increments. In contrast, the control experiments did not show such dynamics. The temperature difference from control in which no freeze-thaw was done at the point of thawing is about 0.05oC. However, if the water being measured is kept in a thermal flask to prevent heat exchange with the local environment, no dropping of temperature were observed under freeze-thaw treatment of the remote water. Statistical analysis performed on data collected after freezing for 10 minutes show that the results are significantly different under these different treatments/settings.

In addition, Figure 4A shows one particular example of temperature variations under remote manipulation of water quantum-entangled with water being measured. In this case, the temperature difference from control at the point of thawing is about 0.08oC. Further, Figure 4B shows one example of temperature variation of a different liquid, Coke Cola, under remote manipulation of a portion of the said liquid quantum-entangled with another portion being measured. Other liquids such as distilled water, olive oil and alcohol also showed similar qualitative results under the same freeze-thaw treatment. Furthermore, preliminary experiments conducted with the temperature measurement done at a location more than 50 miles way from the Dewar also show results similar to those obtained at distances of 50 and 500 feet respectively.

Figure 5 shows weight variations of the first-reservation during the three stages of manipulation of the water in the remote third-reservoir. Before the submersion of the remote third-reservoir into liquid nitrogen the weight being measured drifted lower very slowly. But almost immediately after the remote third-reservoir was submerged into liquid nitrogen, during which the temperature and physical properties of water being manipulated drastically changed, the weight of the first-reservoir dropped at an increased rate, and after the frozen water was taken out the liquid nitrogen and thawed in warm water the weight of the same first stopped dropping and, in some cases, even rose before resuming drifting lower as further discussed below. In contrast, the control experiments did not show such dynamics. The weight difference from control in which no freeze-thaw was done at the point of thawing is about 2.5mg. Statistical analysis performed on data collected after freezing for 10 minutes show that the results are significantly different under these different treatments/settings.

As shown in Figure 5A, in some cases, the weight of the water being measured not only stopped dropping for several minutes but also rose. The signatures of freezing induced weight decreases and thawing induced weight increases for three different thawing times are very clear. In addition, Figure 5B shows one example of weight and temperature variations under the same remote manipulation of water quantum-entangled with water being weighed and measured respectively. Again, the signatures of freezing and thawing induced weight and temperature decreases and increases are respectively very clear. Further, Figure 5C shows another example of weight and temperature variations under another same remote manipulation in which the Dewar was located about 500 feet away from where the measurements were taken. The general background trend of decreasing temperature was due to environmental temperature change. Yet again, the signatures of freezing and thawing induced weight and temperature variations were respectively are very clear. Also, when the remote water was quickly heated to boiling on a stove instead of being frozen in liquid nitrogen, a brief rise of weight in the range of about 0.5mg were repeated observed in several experiments conducted so far.

Furthermore, when the remote water was poured into the 200-gallon water tank instead of being frozen in liquid nitrogen, small but noticeably increased weight losses were repeatedly observed in the several experiments conducted to date. More specifically, before mixing of the water in the remote third-reservoir with water in the water tank the measured weight drifted lower very slowly, but within short time measured in minutes after the water in the remote third-reservoir was poured into the water tank, during which the water in the said tank got quantum-entangled with the water in the third-reservoir, the weight of the first-reservoir dropped at small but increased rate for a period of time. In contrast, the control experiments did not show such dynamics.

Figure 6 shows an example of temperature variations under the respective treatments of adding 5ml concentrated HCl (30%) to the third reservoir, then adding 20g NaOH to the same and third heating the same to boiling point. The signatures of these remote treatments induced temperature changes were clear and repeatedly observable in quite a few experiments conducted to date.

Figure 7 shows the variations of pH value of the water in the first reservoir in experiments done with the setup in Figure 2. As shown, in about 30 minutes after the second-reservoir containing 100ml concentrated HCl (30% by weight) was placed behind the first-reservoir at a distance of 500cm and on the path of the laser beam, during which the water in the first-reservoir got quantum-entangled with the content in the second reservoir, the pH value of the water in the first-reservoir steadily decreased. In contrast, the control experiments did not show such dynamics. Also, the 50mW red laser did not affect the temperature of the water in the first reservoir significantly during the whole treatment. The difference in pH value from control in which HCl was absence is about 0.070 after 50 minutes of exposure to HCl. Statistical analysis performed on data collected after exposure to HCl for 30 minutes show that the results are significantly different from control. Various experiments done with direct additions of HCl to the remote water also repeated showed decreases in pH value in the water being measured.

## 3. DISCUSSIONS & CONCLUSIONS

With all experimental setups and their variations described herein, we have observed clear and reproducible non-local effects with the aids of high-precision analytical instruments and under well-controlled conditions. The physical observables used for measuring the non-local effects are simple ones which can be measured with high precisions. These effects are, even under the most stringent statistical analysis, significantly above and beyond what were noticeable in the control experiments.

Through careful analysis, we have excluded the possibility that the observed weight variation was a secondary local effect due to heat loss and/or sensitivity drift of balance associated with temperature change induced by the remote manipulation. First, during the period of remote manipulation the total temperature change was less than 0.08ºC so the total heat loss for the 175ml water in the first-reservoir is about 60J. In contrast, the weight loss during remote manipulation was on average about 2mg which is $18 \times 10^9$J in energy unit. Second, the first-reservoir and the pan of the balance were separated by 1-inch white foam to prevent heat transfer to the analytic balance. Even in the highly unlikely scenario that this temperature change somehow affected the overall temperature of the balance, the associated sensitivity drift of the balance was about 0.03mg which is 10 times smaller than what's actually observed. In addition, Figures 5A, 5B and 5C also show several other signatures of remote freeze-thaw treatment as the sole cause of the observed weight variations. Therefore, the observed gravity variation is a genuine and direct non-local effect associated with quantum entanglement. However, as with many other important new results, replications by others are the key to independently confirm our results reported here.

We chose to use liquid nitrogen in a large Dewar placed at a distant location for manipulating water in our experiments because it can provide drastic changes in temperature and properties of water in a very short period of time. Our expectation was that, if the quantum entities inside the water being measured are able to sense the changes experienced by the quantum entities in the water being manipulated through quantum entanglement and further utilize the information associated with the said changes, the chemical, thermal and gravitational properties of the water might be affected through quantum entanglement mediated non-local processes (4, 5). The most logical explanation for these observed non-local effects is that they are the consequences of non-local processes mediated by quantum entanglement between quantum entities in the water being measured and the remote water being manipulated as more specifically illustrated below.

First, when pH value of the water in the manipulation reservoir is high or low or is changing under direct manipulation such as extreme cooling or heating or addition of acidic or alkaline chemical, the measured pH in the detecting reservoir shifts in the same direction under the non-local influence of the water in the manipulation reservoir mediated through quantum entanglement and, under the condition that the detecting reserve is able to exchange energy with its local environment, as if H+ in the latter is directly available to water in the detecting reservoir.

Second, when the temperature in the manipulation reservoir is extremely low or high or is changing under direct manipulation such as extreme cooling or heating or addition of heat-generating and/or property-changing chemical such as concentrated HCl or NaOH powder, the temperature in the detecting reservoir changes in the same direction under non-local influence of the water in the manipulation reservoir mediated through quantum entanglement and, under the condition that the detecting reserve is able to exchange heat with its local environment so that the local thermodynamic energy is conserved, as if the heat or lack of it in manipulation reservoir is directly available to the water in the detecting reservoir.

Third, when water in manipulation reservoir is manipulated though extreme cooling, heating or mixing with large quantum-entangled mass, e.g., water, such that the quantum entanglement of the water under manipulation with its local environment changes, the weight of the water in the detecting reservoir also changes under the non-local influence of the manipulation reservoir mediated through quantum entanglement so that, it is hereby predicted, that the global gravitational energy/potential is conserved.

We suggest here that the said quantum entities inside water are likely nuclear spins for the reasons discussed below. Water contains vast numbers of nuclear spins carried by 1H. These spins form complex intra- and inter-molecular networks through various intra-molecular J- and dipolar couplings and both short- and long-range intermolecular dipolar couplings. Further, nuclear spins have relatively long relaxation times after excitations (8). Thus, when a nematic liquid crystal is irradiated with multi-frequency pulse magnetic fields, its 1H spins can form long-lived intra-molecular quantum coherence with entanglement for information storage (9). Long-lived (~ .05 ms) entanglement of two macroscopic electron spin ensembles in room temperature has also been achieved1. Furthermore, spin is a fundamental quantum process and was shown to be responsible for the quantum effects in both Hestenes and Bohmian quantum mechanics (10, 11). Thus, we suggest that quantum-entangled nuclear spins and/or electron spins are likely the mediators of all observed non-local effects reported here (4).

Several important conclusions and implications can be drawn from our findings. First, we have realized non-local signaling using three different physical observables, pH value, temperature and gravity. Second, we have shown that the temperature of water in a detecting reservoir quantum-entangled with water in a remote reservoir can change against the temperature of its local environment when the latter is manipulated under the condition that the water the detecting reservoir is able to exchange heat with its local environment. Third, we have also shown that the gravity of water in a detecting reservoir quantum-entangled with water in a remote reservoir can change against the gravity of its local environment when the latter was remotely manipulated. Our findings imply that the properties of all matters can be affected non-locally through quantum entanglement mediated processes. Second, the second law of thermodynamics may not hold when two quantum-entangled systems together with their respective local environments are considered as two isolated systems and one of them is manipulated. Third, gravity has a non-local aspect associated with quantum entanglement thus can be non-locally manipulated through quantum entanglement mediated processes. Fourth, in quantum-entangled systems such as biological systems, quantum information may drive such systems to a more ordered state against the disorderly effect of environmental heat.

On a more fundamental level, our findings shed new lights on the nature and characteristics of quantum entanglement and gravity, challenge the applicability of Einstein's relativity theories in the quantum domain, provide vital clues for resolution of the measurement problem in quantum mechanics, and support non-local hidden variable based theories such as Bohmian mechanics and a non-local cosmology.

Finally, with respect applications, our findings enable various quantum entanglement assisted technologies be developed. Some of these technologies can be used to manipulate and/or affect remotely various physical, chemical and/or biological systems including human bodies. Other such technologies can be used for non-local signalling and communications between remote locations of arbitrary distances in various ways. Potentially, some of these technologies can also be used to engineer the gravitational properties of physical matters and develop new types of space vehicles.

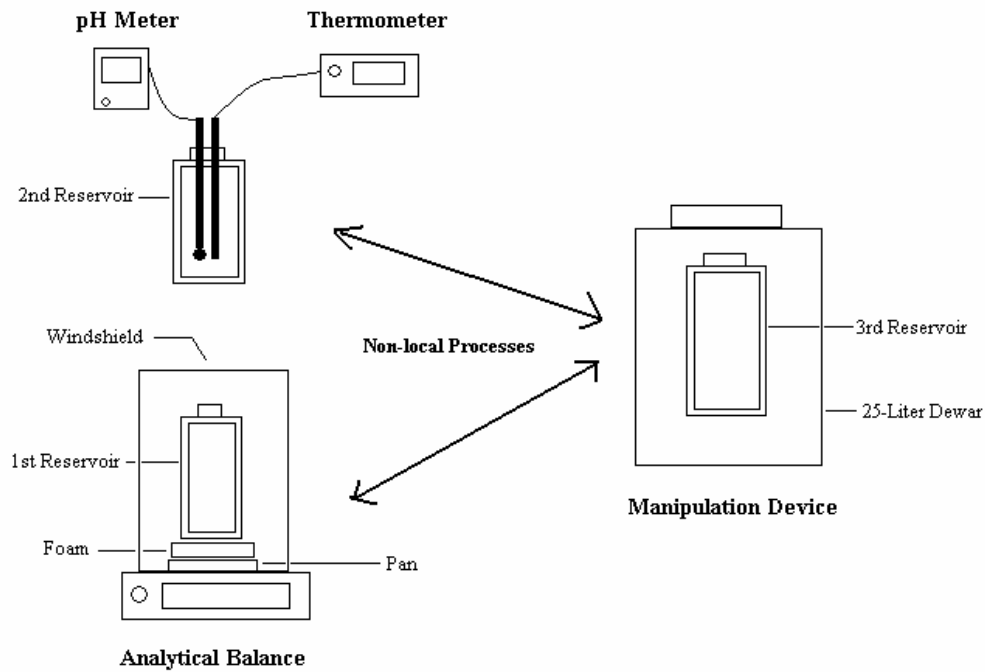

Figure1: Illustration of the key experimental setup. Several variations of this setup were also used in the actual experiments as described in the text. For example, in one variation, the manipulation was heating the water in the 3$^{rd}$ reservoir to boiling point and then cooling it down. In a second variation, the gravity measurement was eliminated and the manipulations were first adding 5*ml* concentrated HCl (30%) to the third reservoir, then adding 20*g* NaOH to the same and third heating the same to boiling point. In a third variation, the Dewar was located more than 500 feet away from the site of measurement. In fourth variation, the gravity and pH measurements were eliminated and the temperature measurements were carried out more than 50 miles away from the location of the Dewar.

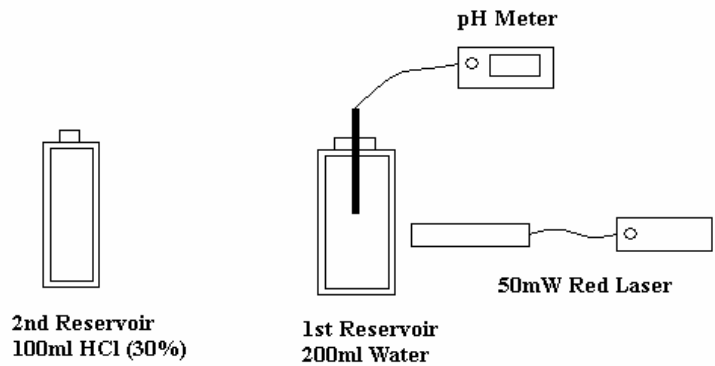

Figure2: Illustration of the second experimental setup which allows the measurement of pH value in the presence or absence of concentrated HCl about 500*cm* away from and behind the water being measured. If no quantum entanglement is involved, the presence or absence of the HCl should not affect the pH value.

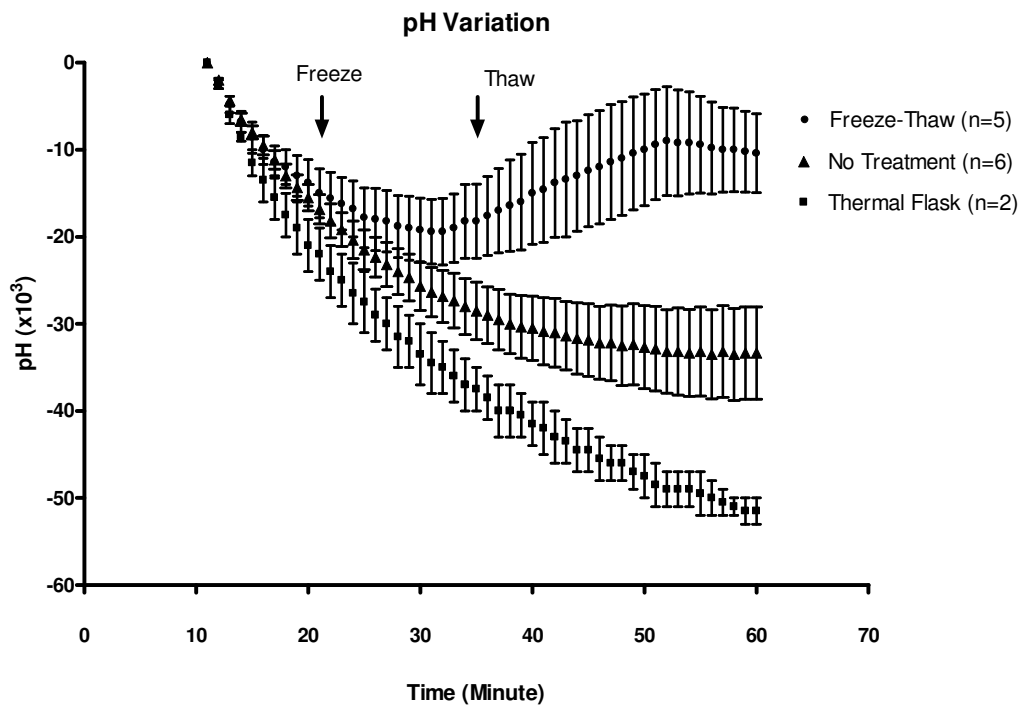

Figure3: pH variations under remote manipulations of water quantum-entangled with water being measured. The pH value at the starting point is set to zero and the results shown were obtained from one batch of quantum-entangled water. The difference in pH values from control in which no freeze-thaw was done at the point of thawing is about 0.010. However, if the water being measured was kept in a thermal flask to prevent energy exchange with the local environment, no effect on pH value was observed during freeze-thaw treatment of remote water. Statistical analysis on data collected after freezing for 10 minutes show that the results are significantly different under the different treatments/settings shown.

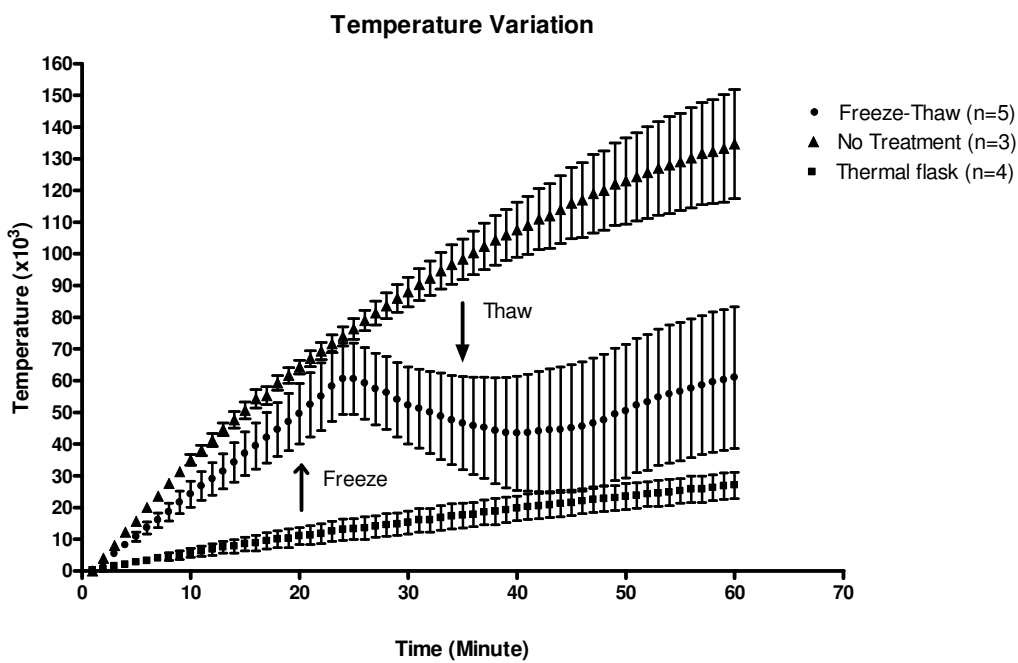

Figure4: Temperature variations under remote manipulations of water quantum-entangled with water being measured. The temperature at the starting point is set to zero and the results shown were obtained from one batch of quantum-entangled water. The temperature difference from control in which no freeze-thaw was done at the point of thawing is about $0.05^{o}C$. However, if the water being measured is kept in a thermal flask to prevent heat exchange with the local environment, no dropping of temperature were observed under freeze-thaw treatment. Statistical analysis performed on data collected after freezing for 10 minutes show that the results are significantly different under the different treatments/settings shown.

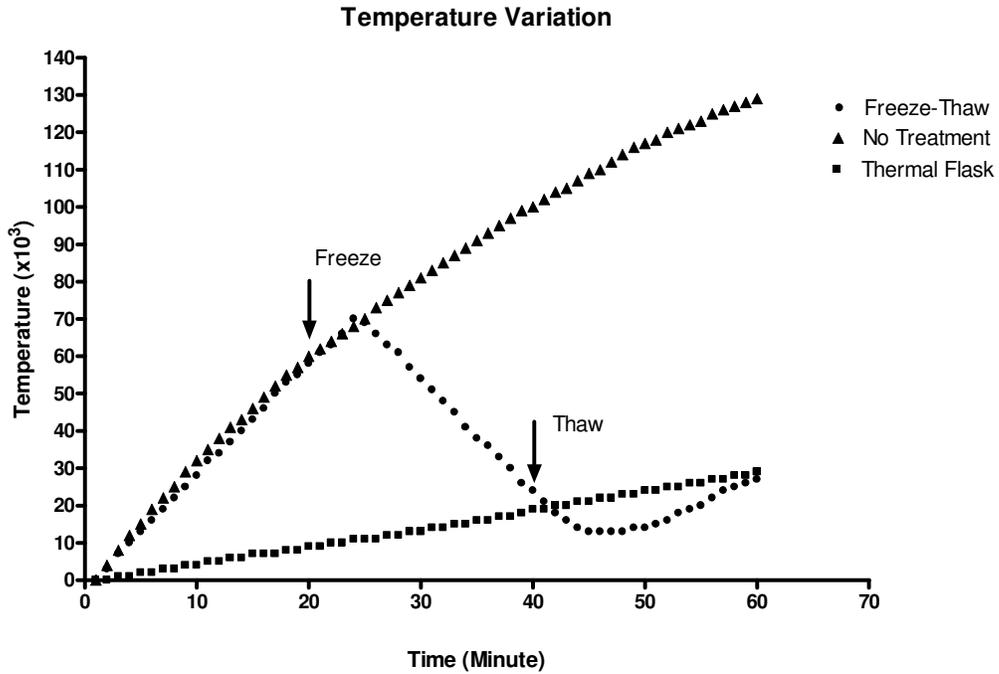

Figure4A: One particular example detailing temperature variations under remote manipulation. The temperature difference from control at the point of thawing is about 0.08°C. However, if the water being measured is kept in a thermal flask, no dropping of temperature were observed under freeze-thaw treatment.

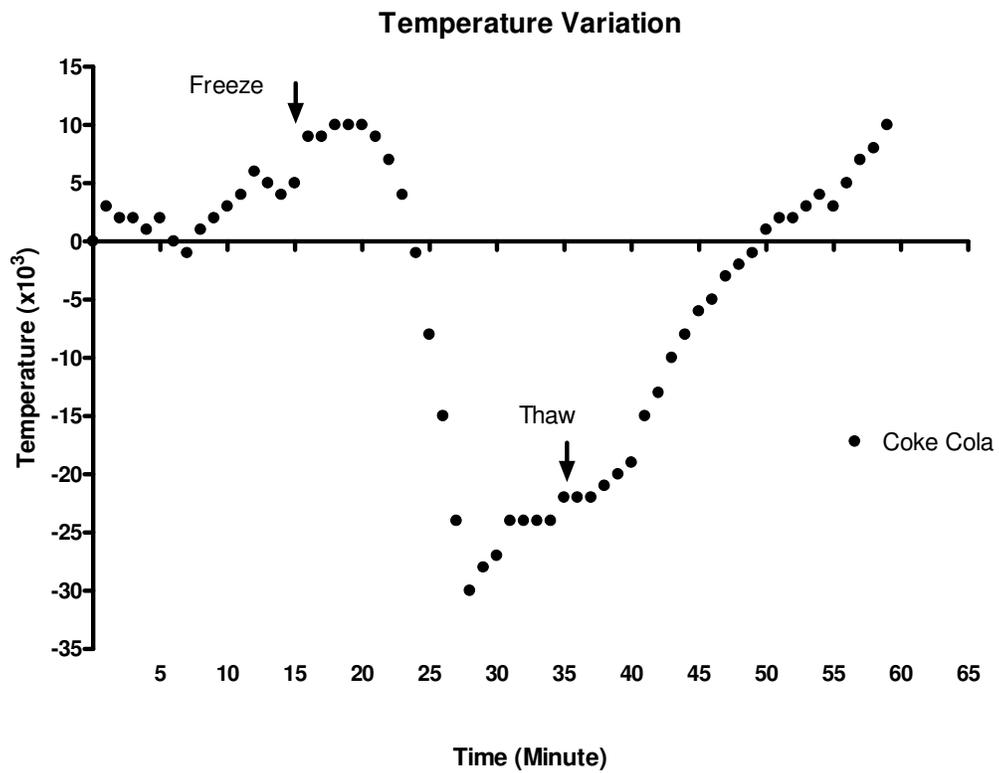

Figure4B: One example showing temperature variation of a different liquid, Coke Cola, under remote manipulation of a portion of the said liquid quantum-entangled with another portion of the liquid being measured. Other liquids such as distilled water, olive oil and alcohol also showed similar qualitative results under the same treatment.

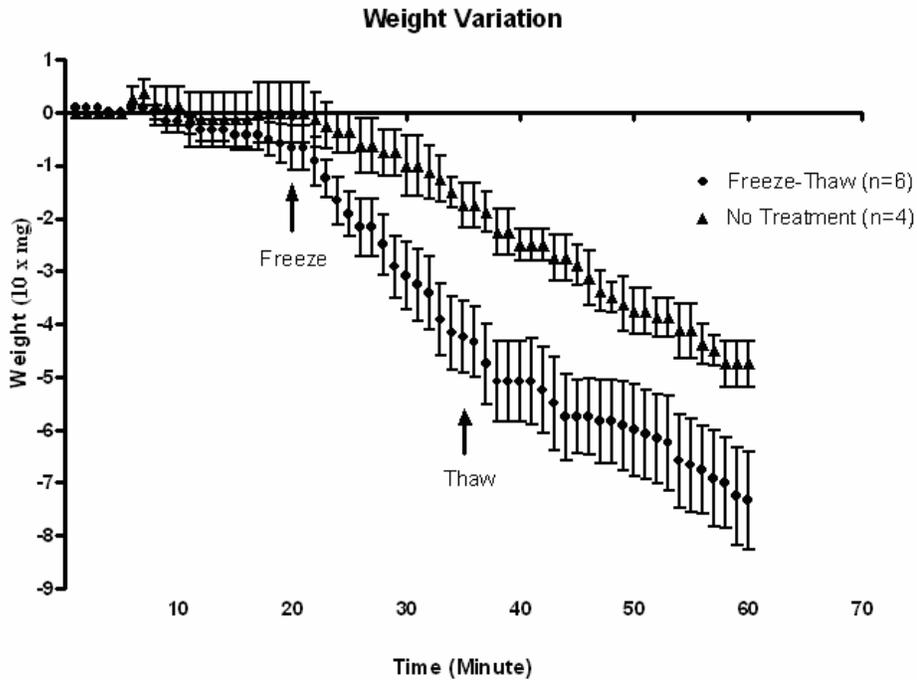

Figure5: Weight variations under remote manipulations of water quantum-entangled with water being weighed. The weight at the starting point is set to zero and the results shown were obtained from one batch of quantum-entangled water. The weight differences from control in which no freeze-thaw was done at the point of thawing is about 2.5*mg*. In some cases, the weight of the water being weighed not only briefly stop dropping for several minutes but also rose briefly for several seconds to minutes as shown in Figure5A. Also when the remote water was quickly heated to boiling on a stove instead of being frozen in liquid nitrogen, a brief rise of weight in the range of about 0.5*mg* were repeated observed in one variation of the key setup. Further, when the remote water was poured into a 200-gallon water tank, small but noticeably increased weight losses were also observed in several experiments conducted to date. Statistical analysis performed on data collected after freezing for 10 minutes show that the results are significantly different under the different treatments/settings shown.

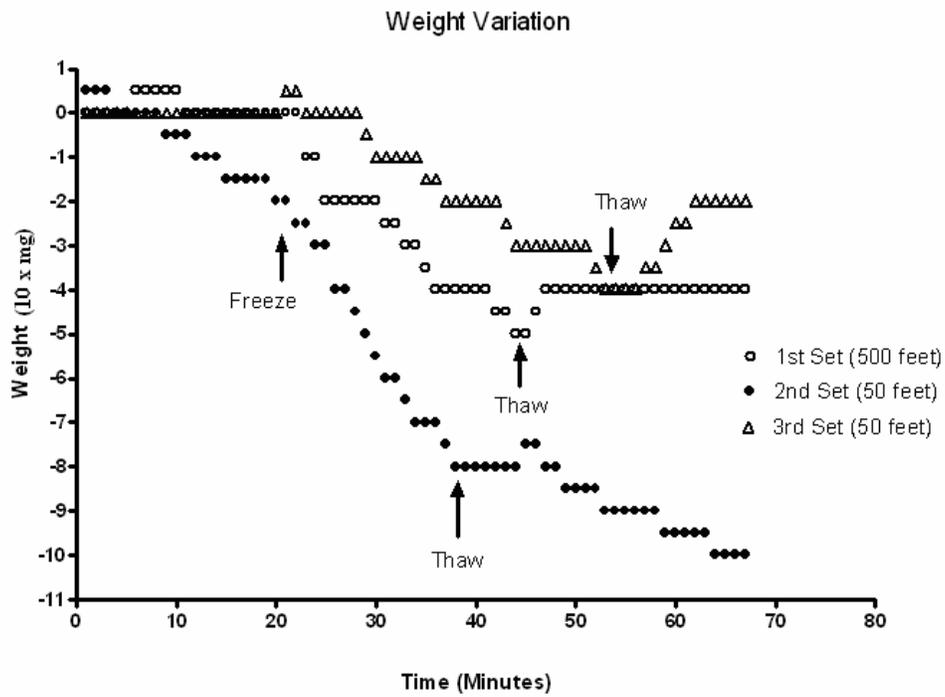

Figure5A: Examples of weight variations under remote manipulations of water quantum-entangled with water being weighed. The onset of increased weight loss started either at the time of freezing treatment or slightly later. The signatures of thawing induced weight increases were clear for the three different thawing times. The distances shown are the respectively distances of the Dewar to the location of measurement in each experiment.

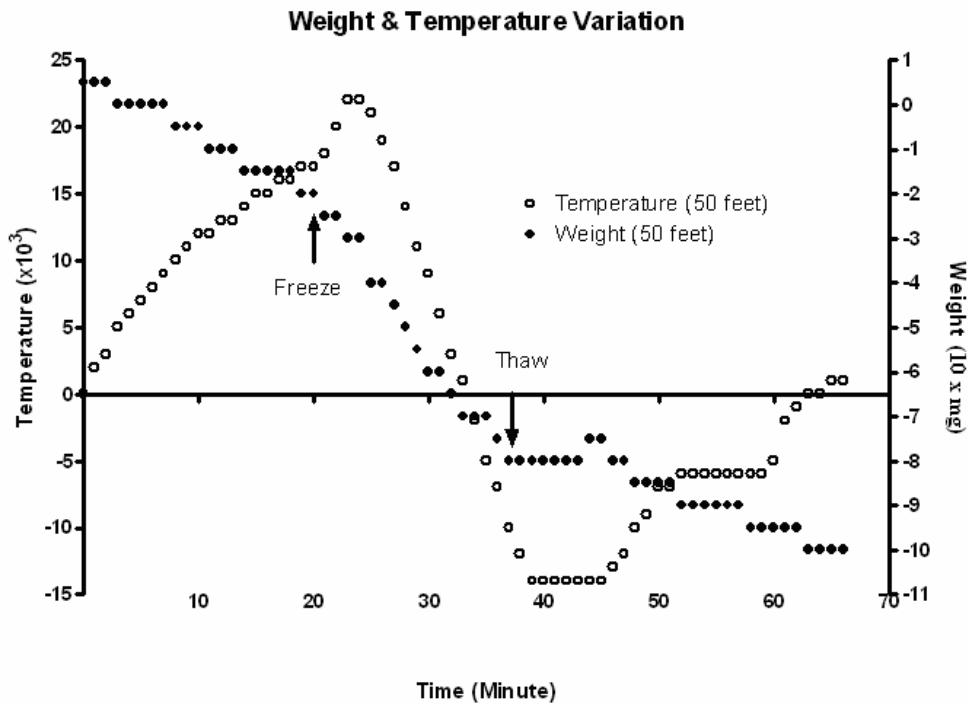

Figure5B: One example of weight and temperature variations under the same remote manipulation of water quantum-entangled with water being weighed and measured respectively. The onset of increased weight loss started at the time of freezing treatment but the onset of temperature decrease against environmental temperature started a few minutes later after freezing treatment started. The signatures of thawing induced weight and temperature increases were clear. The distance shown is the distance of the Dewar to the location of measurement.

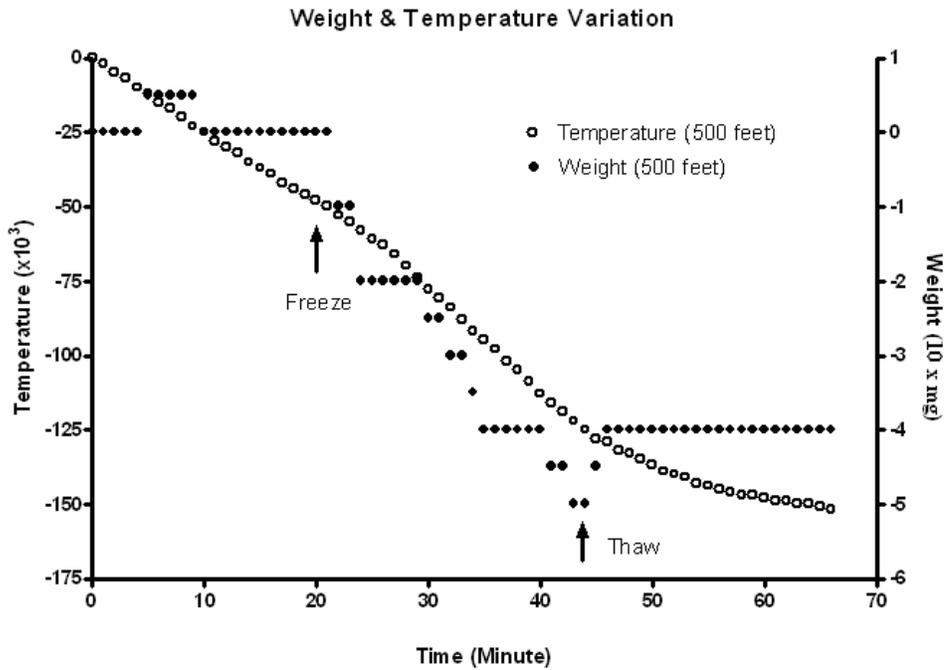

Figure5C: Second example of weight and temperature variations under another same remote manipulation of water quantum-entangled with water being weighed and measured respectively. The general background trend of decreasing temperature was due to environmental temperature change. The onset of increased weight loss started at the time of freezing treatment but the onset of increased temperature loss started a few minutes later after freezing treatment started. The signatures of thawing at time=45*min* induced weight increase and slow down of temperature loss were again clear. The distance shown is the distance of the Dewar to the location of measurement.

**Temperature Variation**

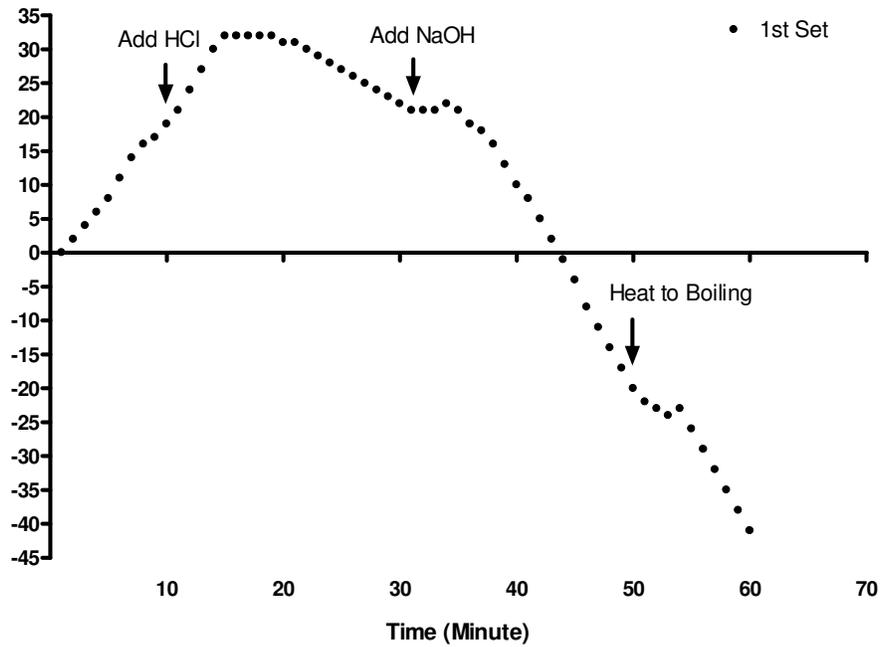

Figure6: An example of temperature variations under the respective treatments of adding 5*ml* concentrated HCl (30%) to the third reservoir, then adding 20*g* NaOH to the same and third heating the same to boiling point. The signatures of these remote treatments induced temperature changes were clear and repeatedly observable in quite a few experiments conducted to date. The general background trend of the temperature first increasing, flattening and decreasing was due to environmental temperature change.

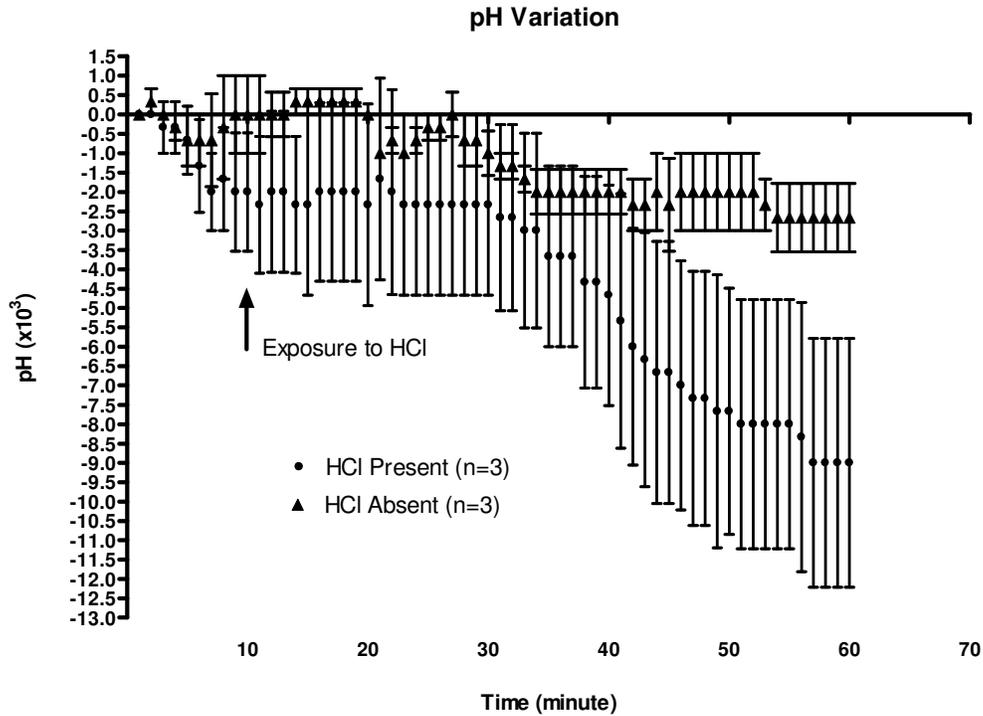

Figure7: pH variations under laser treatment in the presence and absence of concentrated HCl with the setup in Figure 2. The pH value at the starting point is set to zero. The difference in pH value from control in which HCl was absence is about 0.07 after 50 minute of exposure to HCl. Various experiments done with direct additions of HCl to the remote water also repeated showed decreases in pH value in the water being measured. Statistical analysis performed on data collected after exposure to HCl for 30 minutes show that the results are significant different from control.

**ACKNOWLEDGEMENT:** We wish to thank Yongchang Hu and Cuifang Sun for their participation in the experiments Robert N. Boyd for our visit to his place of research. We wish to thank Yongchang Hu for assisting the authors with some of the more recent experiments and Danielle Graham for showing her research at a 2006 conference.

# REFERENCES

## PART I

**PART II**

**PART III**

# APPENDIX I

# THINKING OUTSIDE THE BOX I: THE ESSENCE AND IMPLICATIONS OF QUANTUM ENTANGLEMENT

Huping Hu[1] and Maoxin Wu
(Dated: October 1, 2005)

## ABSTRACT

Many experiments have shown that quantum entanglement is physically real. In this paper, we will discuss its ontological origin, implications and applications by thinking outside the standard interpretations of quantum mechanics. We argue that quantum entanglement originates from the primordial spin processes in non-spatial and non-temporal pre-spacetime, implies genuine interconnectedness and inseparableness of once interacting quantum entities, plays vital roles in biology and consciousness and, once better understood and harnessed, has far-reaching consequences and applications in many fields such as medicine and neuroscience. We further argue that quantum computation power also originates from the primordial spin processes in pre-spacetime. Finally, we discuss the roles of quantum entanglement in spin-mediated consciousness theory.

**Key Words:** Spin, Entanglement, Interconnectedness, Inseparableness

## 1.      Introduction

Quantum entanglement is ubiquitous, appears everywhere in the microscopic world (See, e.g., Durt, 2004; Brooks, 2005) and under some circumstances manifests itself macroscopically (Arnesen, *et al*, 2001; Ghost *et al*, 2003 & Julsgaard *et al*, 2001). Indeed, it is currently the most intensely studied subject in physics. Further, speculations abound as to its nature and implications (See, e.g., Clarke, 2004, Josephson, 1991 & Radin, 2004). There are many general and technical papers written on the subject. So cutting to the chase, we shall immediately outline our propositions on the subject and then discuss each in some detail with references to existing literature whenever possible. Readers are advised that our propositions are outside the mainstream physics and other authors may hold similar views on some of the points we shall make in this paper. We will also discuss the roles of quantum entanglement in spin-mediated consciousness theory (Hu & Wu, 2002, 2003, 2004a-d).

---

[1]Corresponding author: Huping Hu, Ph.D., J.D., Biophysics Consulting Group, 25 Lubber Street, Stony Brook, NY 11790, USA. E-mail: hupinghu@quantumbrain.org

The following are our propositions about the ontological origin, implications and applications of quantum entanglement besides quantum computation:

1) It originates from the primordial spin processes in non-spatial and non-temporal pre-spacetime. It is the quantum "glue" holding once interacting quantum entities together in pre-spacetime, implies genuine interconnectedness and inseparableness of the said quantum entities and can be directly sensed and utilized by the entangled quantum entities.
2) Thus, it can influence chemical/biochemical reactions, other physical processes and micro- and macroscopic properties of all forms of matters, thus, playing vital roles in many biological processes and consciousness. It is the genuine cause of many anomalous effects (if they do exist) in parapsychology, alternative medicine and other fields as some authors have already suspected in some cases.
3) Further, it can be harnessed, tamed and developed into revolutionary technologies to serve the mankind in many areas such as health, medicine and even recreation besides the already emerging fields of quantum computation.

## 2. The Origin and Nature of Quantum Entanglement

Popular opinion has it that Erwin Shrödinger coined the word "entanglement" and first used it in 1935 in his article published in the Proceedings of Cambridge Philosophical Society (Shrödinger, 1935). Mathematically, Shrödinger showed that entanglement arises from the interactions of two particles through the evolution of Shrödinger equation and called this phenomenon the characteristic trait of quantum theory (*id.*). Einstein called quantum entanglement "spooky action at a distance" in the famous EPR debate (See, e.g., Einstein *et al*, 1935).

Ontologically, we argue that quantum entanglement arises from the primordial self-referential spin processes which we had argued previously are the driving force behind quantum effects, spacetime dynamics and consciousness (Hu & Wu, 2003; 2004a). Pictorially, two interacting quantum entities such as two electrons get entangled with each other through the said spin processes by exchanging one or more entangling photons with entangling occurring in pre-spacetime. Such ontological interpretation is supported by existing literature as discussed below.

First, Hestenes showed that in the geometric picture for the Dirac electron the zitterbewegung associated with the spin is responsible for all known quantum effects of said electron and the imagery number $i$ in the Dirac equation is said to be due to electronic spin (See, e.g., Hestines, 1983).

Second, in Bohmian mechanics the "quantum potential" is responsible for quantum effects (Bohm and Hiley, 1993). Salesi and Recami (1998) have recently shown that said potential is a pure consequence of "internal motion" associated with spin evidencing that the quantum behavior is a direct consequence of the fundamental existence of spin. Esposito (1999) has expanded this result by showing that "internal motion" is due to the spin of the particle, whatever its value. Bogan (2002) has further expanded these results by deriving a spin-dependent gauge transformation between the Hamilton-Jacobi equation of classical mechanics and the time-dependent Shrödinger equation of quantum mechanics which is a function of the quantum potential of Bohmian mechanics.

Third, spin is a unique quantum concept often being said to have no classical counterpart (See Tomonaga, 1997) and associated with the "internal motion" of a point particle. Unlike mass and charge that enter a dynamic equation as arbitrary parameters, spin reveals itself through the structure of the relativistic quantum equation for fermions that combines quantum mechanics with special relativity (Dirac, 1928). Indeed, many models of elementary particles and even space-time itself are built with spinors (Budinich, 2001; Penrose, 1960 & 1967). Pauli (1927) and Dirac (1928) were the first to use spinors to describe the electron. Also, Kiehn (1999) showed that the absolute square of the wave function could be interpreted as vorticity distribution of a viscous compressible fluid that also indicates that spin is the process driving quantum mechanics.

Therefore, in view of the foregoing it could be said that the driving force behind the evolution of Shrödinger equation is quantum spin and, since quantum entanglement arises from the evolution of Shrödinger equation the said spin is the genuine cause of quantum entanglement.

What do we mean by pre-spacetime? Pre-spacetime in this article means a non-spatial and non-temporal domain but it is not associated with an extra-dimension in the usual sense since there is no distance or time in such domain (See, *e.g.*, Hu & Wu, 2002). We have argued before that in a dualistic approach mind resides in this domain and unpaired nuclear and/or electronic spins are its pixels in the reductionist perspective (*id.*). So pre-spacetime is a holistic domain located outside spacetime but connected through quantum thread/channel to everywhere in spacetime enabling quantum entanglement or Einstein's "spooky action at a distance." It has similarity to Bohm's concept of implicate order (Bohm & Hiley, 1993). Aerts (2004), Clarke (2004) and others have also expressed the non-space view of quantum nonlocality.

Such a non-spatial and non-temporal pre-spacetime is a "world" beyond Einstein's relativistic world but does not contradict with the latter since the latter deals with classical physical events occurring within spacetime. In contrast, quantum entanglement occurs within non-spatial and non-temporal domain. Therefore, instantaneous signaling through quantum entanglement in pre-spacetime is possible if the entangled quantum entities can directly sense and/or utilize the entanglement.

So what is then the essence of quantum entanglement? We propose that quantum entanglement is not merely the correlations of certain observable physical parameters in the process of measurement but genuine interconnectedness and inseparableness of once interacting quantum entities. It is the quantum "glue" holding once interacting quantum entities together in pre-spacetime and can be directly sensed and utilized by the entangled quantum entities as further discussed below. It can be diluted through entanglement with the environment, *i.e.*, decoherence.

### 3. Implications of Quantum Entanglement

It is often said that instantaneous signaling through quantum entanglement is impossible because of Eberhard's theorem that basically says that since local measurements always produce random results no information can be sent through quantum entanglement alone (Eberhard, 1978). However, there are at least two ways to circumvent this impossibility. The first is to assume that the statistical distribution can be modified and utilized to transmit information. Quite a few authors have expressed this view (Josephson, 1991; Stapp, 1982 & Walker, 1974) especially when discussing the roles of consciousness in parapsychology such as telepathy. The second is to assume that each quantum entity can directly sense and utilize quantum entanglement as already mentioned before. This latter view is the view we subscribe to and it is permissible in the Bohmian picture (Bohm & Hiley, 1993).

The implication of the second view is far-reaching. It means that quantum entanglement can influence chemical and biochemical reactions and other physical processes. Thus, it plays vital roles in many biological processes and consciousness and is the genuine cause of many anomalous effects, if they do exist, in parapsychology, alternative medicine and other fields as some authors have already suspected in some case. It can affect the micro- and macroscopic properties of all forms of matters such solid and liquid.

For example, the results reported by Rey (2003) that heavy water and highly diluted solutions of sodium chloride and lithium chloride behaved differently in the thermo- luminescence tests can be explained as the consequence of water molecules forming different hydrogen bonds due to the entanglement of water molecule with sodium chloride or lithium chloride ions being diluted out of existence and its subsequent effect on hydrogen bond formation during freezing. Indeed, in light of the recent results on observable macroscopic entanglement effects (Arnesen, *et al*, 2001; Ghost *et al*, 2003), the explanation offered herein is most likely true.

For a second example, the so called "memory of water" effect (Davenas, *et al*, 1988), which is largely discredited by the mainstream scientists because of non-reproducibility, can be explained as the result of entanglement of the substances being diluted with water and then the subsequent entanglement of water with the quantum entities in the biochemical processes responsible for producing certain visible or detectable result. Of course, quantum entanglement cannot directly serve as a reagent in a chemical reaction nor can it be recorded or transferred through any classical means such as a digital device within a computer or the telephone wire. So any claim of recordable or telephone -wire-transferable "chemical signal" cannot be attributed to quantum entanglement.

Similarly, the therapeutic effect of a homeopathic remedy, if it truly exists beyond and above the placebo effect, can be explained as the entanglement of the substances being diluted out of existence through vigorous shaking/stirring with the diluting solvent and then the subsequent entanglement of the solvent with the quantum entities involved in the diseased biological and/or physiological processes and the effect of such entanglement on the latter processes. Indeed, there are reports in the existing literature exploring the use of generalized entanglement to explain the therapeutic ingredient in a homeopathic remedy (See, e.g., Milgrom, 2002; Wallach, 2000 & Weingärtner, 2003).

Further, many other unconventional healing effects reported in alternative medicine such as Qi Gong and other types of bioenergy healing, if they are genuine, can be explained as the results of quantum entanglement between the quantum entities involved in the diseased processes and the quantum entities in the healing sources, such as a healthy biological entity, and the effect of the former on the latter processes.

For yet another example, all the results from Princeton Engineering Anomalies Research program over the last 26 years (Jahn & Dunne, 2005) can also be straightforwardly explained as the entanglement of the quantum entities controlled by human mind with the quantum entities responsible for the physical processes capable of producing modified random results. By the same token, many if not all anomalous effects reported in parapsychology such as telepathy and those results reported by Grinberg- Zylberbaum (1987) and the repeaters (For a summary, see, Wackermann, 2005) can be simply explained as the results of quantum entanglement between the quantum entities capable of invoking action potentials in one person and those in a second person and the effect of one on the other through quantum entanglement. Grinberg-Zylberbaum himself speculated that his results had something to do with quantum entanglement (1994).

4.   **The Ontological Origin of Quantum Computation Power**

It is said that the computational speed-up of a quantum computer is due to quantum entanglement (See, *e.g.*, Steane, 2000). However, the ontological origin of its power over classical computation is very much in dispute due to different interpretations of quantum mechanics (*id.*).

For example, some argue that, in terms of the amount of information manipulated in a given time, quantum superposition/entanglement permits quantum computers to ``perform many computations simultaneously'' which invoke the concept of vast numbers of parallel universes (See, *e.g.*, Deutsch & Hayden, 2000; Deutsch, 2002). Others argue that quantum entanglement makes available types of computation process, which, while not exponentially larger than classical ones, are unavailable to classical systems (See, *e.g.*, Steane, 2000). Thus, according to Steane (2000), the essence of quantum computation is that it uses entanglement to generate and manipulate a physical representation of the correlations between logical entities, without the need to completely represent the logical entities themselves (*id.*).

Do we have anything to add in light of our view expressed in this paper? The answer is "Yes." We argue that the types of computation process made available by quantum entanglement are the ones driven by the primordial spin processes in the non-spatial and non-temporal pre-sapcetime. Or, if you like, the power of quantum computation over classical computation originates from Bohm's implicate order driven by the primordial spin processes.

**5.    Applications of Quantum Entanglement**

Recently, quantum computations have been achieved in the laboratory but they are implemented in controlled environment to prevent decoherence through entanglement of the system of interests with its surrounding environment. Indeed, it is also often said that the reason why we don't experience quantum entanglement in the macroscopic world is because of rapid decoherence within the macroscopic system. However, this view may rapidly change (See, *e.g.*, Brooks, 2005). We are convinced that quantum entanglement can be harnessed, tamed and developed into revolutionary technologies to serve the mankind in many areas such as health, medicine and even recreation besides the emerging fields of quantum computation and communications.

For example, once harnessed, quantum entanglement technologies can be used to deliver the therapeutic effects of many drugs to a target biological system such as a human body without ever physically administrating the said drugs to the said system. Such technology would dramatically reduce waste and increase productivity because the same drugs can be repeatedly used to deliver their therapeutic effects to the mass. By the same token, many substances of nutritional and even recreational values can be repeatedly administrated to the human body through the said technologies. For a second example, the harnessed quantum entanglement technologies can also be used to entangle two or more human minds for legitimate purposes. Further, the said technologies can be used for instantaneous communications with humans sent to the outer space.

Are we delusional? We think not. We predict that the wonders of quantum entanglement technologies will be soon widely utilize to serve the mankind and a new paradigm of science will be born in the near future.

## 6. Quantum Entanglement in Spin-Mediated Consciousness Theory

Our spin-mediated consciousness theory says that quantum spin is the seat of consciousness and the linchpin between mind and the brain, that is, spin is the mind-pixel (Hu & Wu, 2002, 2004a-d). The starting point is the fact that spin is basic quantum bit ("qubit") for encoding information and, on the other hand, neural membranes and proteins are saturated with nuclear spin carrying nuclei and form the matrices of brain electrical activities. Indeed, as discussed above, spin is embedded in the microscopic structure of spacetime as reflected by Dirac equation and is likely more fundamental than spacetime itself as implicated by Roger Penrose's work. In the Hestenes picture the zitterbewegung associated with spin was shown to be responsible for the quantum effects of the fermion. Further, in the Bohm picture the internal motion associated with spin has been shown to be responsible for the quantum potential which, in turn, is responsible for quantum effects. Thus, if one adopts the minority quantum mind view, nuclear spins and possibly unpaired electron spins become natural candidates for mind-pixels (Hu & Wu, 2002; 2003; 2004a-d).

Applying these ideas to the particular structures and dynamics of the brain, we have theorized that human brain works as follows: Through action potential modulated nuclear spin interactions and paramagnetic O2/NO driven activations, the nuclear spins inside neural membranes and proteins form various entangled quantum states and, in turn, the collective dynamics of the said entangled quantum states produces consciousness through contextual, irreversible and non-computable means and influences the classical neural activities through spin chemistry (Hu & Wu, 2002; 2003; 2004a-d).

Existing literature supports the possibility of a spin-mediated consciousness. For example, it was shown that proton nuclear spins in nematic liquid crystal could achieve long-lived intra-molecular quantum coherence with entanglement in room temperature for information storage (Khitrin *et al*, 2002). Long-ranged (>10 microns) intermolecular multiple-quantum coherence in NMR spectroscopy was discovered about a decade ago (Warren, et al 1993). Long-lived (>.05 milliseconds) entanglement of two macroscopic spin ensembles in room temperature has been achieved recently (Julsgaard, *et al.* 2001). Further, NMR quantum computation in room temperature is reality (Gershenfeld & Chuang, 1997).

Therefore, according to our theory, consciousness is intrinsically connected to the spin process and emerges from the collective dynamics of various entangled spin states and the unity of mind is achieved by entanglement of these mind-pixels (Hu & Wu, 2002; 2003; 2004a-d). Our theory is tentative as are all current theories about consciousness. As with other quantum mind theories, decoherence is a major concern as pointed out by Tegmark (2000) but may not be insurmountable (See, e.g., Hagan, *et.al.*, 2002). We believe that the solution will be found through the study of the nature of quantum entanglement.

Indeed, our dualistic approach adopted earlier (Hu & Wu, 2002) and described in more detail in this paper allows mind to utilize quantum entanglement to achieve the unity of mind in pre-spacetime. The essential question is then how does mind process and harness the information from the mind-pixels which form various entangled spin states so that it can have conscious experience. We have argued that contextual, irreversible and non-computable means within pre-spacetime are utilized by mind to do this. Furthermore, we recognize that there may not be any large-scale quantum coherence in the warm and wet brain to support those quantum theories of mind that require macroscopic quantum effects. However, our theory does not depend on such a coherence to work in the dualistic approach.

## 7. Conclusion

In this article, we have discussed the ontological origin, implications and applications of quantum entanglement by thinking outside the standard interpretations of quantum mechanics. We have argued that quantum entanglement originates from the primordial self-referential spin processes in non-spatial and non-temporal pre-spacetime, implies genuine interconnectedness and inseparableness of once interacting quantum entities, play vital roles in biology and consciousness and, once better understood and harnessed, has far-reaching consequences and applications in many fields such as medicine and neuroscience. It follows then that quantum computation power also originates from the primordial spin processes in pre-spacetime. We have also discussed the roles of quantum entanglement in our spin-mediated consciousness theory.

Finally, the principle of science dictates that a scientific theory/hypothesis should only achieve legitimacy if it is experimentally verified. Thus, since the summer of 2004 to the present, we have mainly focused our efforts on the quantification of our theory and the designs and implementations of computer simulations and experiments for the verifications of the same. Important results shall be reported as soon as feasible.

# APPENDIX II

# THINKING OUTSIDE THE BOX II: THE ORIGIN, IMPLICATIONS AND APPLICATIONS OF GRAVITY AND ITS ROLE IN CONSCIOUSNESS

Huping Hu[3] and Maoxin Wu

(Dated: 11/16/2006)

## ABSTRACT

Although theories and speculations abound, there is no consensus on the origin or cause of gravity. Presumably, this status of affair is due to the lack of any experimental guidance. In this paper, we will discuss its ontological origin, implications and potential applications by thinking outside the mainstream notions of general relativity and quantum gravity. We argue that gravity originates from the primordial spin processes in non-spatial and non-temporal pre-spacetime, is the manifestation of quantum entanglement, and implies genuine instantaneous interconnectedness of all matters in the universe. That is, we advocate the principle of non-local action. To certain degree, our view is a reductionist expression of Newton's instantaneous universal gravity and Mach's Principle with important consequences. We also discuss the role of gravity in consciousness from this new perspective. Indeed, if spin is the primordial self-referential cause of everything, it should also be the cause of gravity.

**Key Words:** Gravity, Spin, Entanglement, Instantaneity, Interconnectedness, Consciousness

## 1. Introduction

This paper is an extension of our earlier papers advocating a holistic and unified theme of reality in which spin is the primordial self-referential process driving quantum mechanics, spacetime dynamics and consciousness (1-5). Briefly, we have proposed a spin-mediated consciousness theory in which spin is the mind-pixel (1), outlined a unified theme of reality based on spin (2), offered our views on the essence and implications of quantum entanglement (3), and conducted experiments which indeed support our propositions (4-5). In particular, our experiments show that (a) biologically/chemically meaningful information can be transmitted through quantum entanglement; (b) both classical and quantum information can be transmitted between locations of arbitrary distances through quantum entanglement; (c) instantaneous signaling is physically real; and (d) brain processes such as perception and other biological processes likely involve quantum information and nuclear and/or electronic spins are likely play important roles in these processes.

[1]Corresponding author: Huping Hu, Ph.D., J.D., Biophysics Consulting Group, 25 Lubber Street, Stony Brook, NY 11790, USA. E-mail: hupinghu@quantumbrain.org

While our primary focus at the present is till experimental studies, we would like to offer our view on the origin, implications and potential applications of gravity. The connection between quantum entanglement and Newton's instantaneous universal gravity and Mach's Principle is natural. To certain degree, our view is a reductionist expression of this connection with important consequences. Readers are again advised that our propositions are outside mainstream physics and other authors may hold similar views on some of the points we shall make. Some readers are further advised to treat this paper as an Ed-Op piece so as to avoid being offended. However, we are deadly serious about our propositions and put our money where our mouth is.

Microscopically gravity is assumed to be fable and negligible and macroscopically it is ubiquitous and pervasive. It seems to penetrate everything and cannot be shielded. However, there is no consensus as to its cause despite of the efforts of many people. Presumably, this status of affair is due to the lack of any experimental guidance. There are many general and technical papers written on the subject. So cutting to the chase, we shall immediately outline our propositions and then discuss each in some detail with reference to existing literatures whenever possible. Our propositions are as follows:

4) Gravity originates from the primordial spin processes in non-spatial and non-temporal pre-spacetime and is the macroscopic manifestation of quantum entanglement.
5) Thus, gravity is nonlocal and instantaneous, as Newton reluctantly assumed and Mach suggested. It implies that all matters in the universe are instantaneously interconnected and many anomalous effects in astronomy such as red shift, dark energy, dark mass and Pioneer effect may be resolved from this perspective.
6) Potentially, gravity can be harnessed, tamed and developed into revolutionary technologies to serve the mankind in many areas such as instantaneous communication, spacetime engineering and space travel.

## 2. The Origin and Nature of Gravity

The idea of instantaneous gravity is nothing new. Newton's law of universal gravitation implies instantaneous "action at a distance" which he felt deeply uncomfortable with, but Newton was not able to find a cause of gravity (6). Later

Mach suggested that "[t]he investigator must feel the need of... knowledge of the immediate connections, say, of the masses of the universe…[t]here will hover before him as an ideal insight into the principles of the whole matter, from which accelerated and inertial motions will result in the same way" (7). Ontologically, Mach's above suggestion is a form of holism and implies that gravity is relational and instantaneous.

It was Einstein who fulfilled Mach's "relational" suggestion of gravity by inventing general relativity (8). He also coined the phrase Mach's principle. However, such fulfillment is at the sacrifice of Mach's "immediate connections" by assuming that the speed of gravity is the speed of light. Einstein's general relativity is now the mainstream theory of gravity, but it is in conflict with quantum mechanics – the most successful theory of the 20$^{th}$ century which Einstein himself helped to build. Einstein called quantum entanglement "spooky action at a distance" in the famous EPR debate (9). However, it seems that Einstein's camp is on the losing side of the debate today as many recent experiments have shown that quantum entanglement is physically real (e.g., 10-11). We suggest that a theory of gravity, which includes general relativity as an approximation, be built from the properties of quantum entanglement.

Ontologically, we have argued that quantum entanglement arises from the primordial self-referential spin processes which are envisioned by us as the driving force behind quantum mechanics, spacetime dynamics and consciousness (2). Pictorially, two interacting quantum entities such as two electrons get entangled with each other through the said spin processes in pre-spacetime. Such ontological interpretation is supported by existing literature as discussed previously (2). Here we focus our discussion on spin as the primordial process driving space-time dynamics including gravity.

First, spin is deeply connected to the microscopic structure of spacetime as reflected by the Dirac equation for Dirac spinor field representing the fermions (12). Indeed, Penrose (13-14) had considered early on that spin might be more fundamental than spacetime and invented spinor and twistor algebras for a combinatorial description of spacetime geometry. Bohm and Hiley (15) generalized the twistor idea to Clifford algebra as a possible basis for describing Bohm's "implicit order." Recently various spin foams have been formulated as extensions to Penrose's spin networks for the purpose of constructing a consistent theory of quantum gravity (16). Many others have also study the nature of spin from both classical and quantum-mechanical perspectives. For example, Newman showed that spin might have a classical geometric origin. By treating the real Maxwell Field and real linearized Einstein equations as being embedded in complex Minkowski space, he was able to interpret spin-angular momentum as arising from a charge and "mass monopole" source moving along a complex world line (17).

Second, Sidharth (18-19) has discussed the nature of spin within the context of quantized fractal spacetime and showed that spin is symptomatic of the non-commutative geometry of space-time at the Compton scale of a fermion and the three dimensionality of the space result from the spinorial behavior of fermions. He showed that mathematically an imaginary shift of the spacetime coordinate in the Compton scale of a fermion introduces spin ½ into general relativity and curvature to the fermion theory (18). The reason why an imaginary shift is associated with spin is to be found in the quantum mechanical zitterbewegung within the Compton scale and the consequent quantized fractal space-time (18). Further, according to Sidharth (19), a fermion is like a Kerr-Newman black hole within the Compton scale of which causality and locality fails.

Third, Burinskii has recently shown that in spite of the weakness of the local gravitational field, the gravity for a spin ½ fermion as derived using the classical Kerr-Newman Kerr solution (Kerr's Gravity) has very strong stringy, topological and non-local action on the Compton distances of the fermion, polarizing the space-time and electromagnetic field and controlling the basic quantum properties of the fermions (20). Thus, Kerr's Gravity may suggest possibly deep connections between the mass-energy relationship of matter and the quantum properties of particles (20).

Fourth, Makhlin has recently shown that the axial field component in the spin connections of the Dirac spinor field provides an effective mechanism of auto-localization of the Dirac spinor field into compact objects, presumably representing the fermions, and condition that the compact objects are stable leads to the Einstein's field equations (21). He suggested that the physical origin of the macroscopic forces of gravity between any two bodies is a trend of the global Dirac spinor field to concentrate around the microscopic domains where this field happens to be extremely localized (21). He further suggested that the long distance effect of the axial field is indistinguishable from the Newton's gravity which according to him reveals the microscopic nature of gravity and the origin of the gravitational mass (21).

Further, Penrose-Hameroff's self-organized objective reduction model of spacetime geometry (22) also implies that the spacetime dynamics is driving by certain self-referential process. In addition, Cahill's work on a self-referentially limited neural-network model of reality (23) supports the view of a primordial self-referential network underlying reality.

We emphasize that pre-spacetime in this article means a non-spatial and non-temporal domain but it is not associated with an extra-dimension in the usual sense since there is no distance or time in such domain (2-3). So pre-spacetime is a holistic domain located outside spacetime but connected through quantum entanglement to everywhere in spacetime enabling Newton's instantaneous universal gravity and Mach's "immediate connections." It has similarity to Bohm's concept of implicate order (24) and other non-local hidden variable theories (25-26). The said pre-spacetime is a "world" beyond Einstein's relativistic world through which quantum entanglement can be used to produce instantaneous signaling as we have demonstrated experimentally (4-5).

In short, existing literatures cited above support the proposition that spin is the primordial process driving space-time dynamics. Since we have argued previously that quantum entanglement also originates from the primordial spin process, it is natural to link gravity to the property of quantum entanglement. Indeed, doing so will not only provide a cause to Newton's instantaneous universal gravity but also realize Mach's "immediate connections' discussed above. Therefore, we propose that gravity originates from the primordial spin processes in non-spatial and non-temporal pre-spacetime and is the macroscopic manifestation of quantum entanglement.

## 3.     Implications of Gravity from This New Perspective

At the present we are contemplating a mathematical framework from which the primordial self-referential spin process produces everything including gravity. There are several exiting approaches which provide some hints as to the said mathematical forms. These approaches are all based non-local hidden variables, that is, the principle of non-local action. They include Bohmian mechanics (15, 24), Adler's trace dynamics (25), Smolin's stochastic approach (26) and Cahill's process physics (23).

In additions, other existing alternative approaches on gravity may also provide some hints. For sample, Sakharov's induced gravity is a well known alternative theory of quantum gravity in which gravity emerges as a property of matter fields (27). In comparison, we advocate herein that gravity is a property of quantum entanglement.

The implication of this new perspective on gravity is far-reaching. It implies that gravity is non-local and instantaneous, as Newton reluctantly assumed and Mach suggested, all matters in the universe are instantaneously interconnected and, therefore, many anomalous effects in astronomy such as dark energy, dark mass and Pioneer effects may be resolved from this perspective and within a framework of non-local cosmology.

## 4.     Potential Applications of Gravity from This New Perspective

We are convinced that gravity, as the macroscopic manifestation of quantum entanglement, can be harnessed, tamed and developed into revolutionary technologies to serve the mankind in many areas.

For example, once harnessed, non-local gravity may be used to communicate between locations of arbitrary distances instantaneously. Such technology would dramatically decrease the communication delay to spacecraft in outer space. For a second example, once tamed, powerful non-local gravity may be used to engineer the structures of spacetime and propel a new kind of spacecraft for advanced space travel.

Is this for real? You bet. We put our money where our mouth is and predict that the wonders of non-local gravity technologies will soon be widely utilize to serve the mankind and a new paradigm of science will be born in the near future.

5.  **The Role of Gravity in Consciousness**

As manefestation of quantum entanglement, the role of gravity in consciousness is to achieve binding. In our spin-mediated consciousness theory, such role played by gravity is not hard to see, since spin is the seat of consciousness and the linchpin between mind and the brain, that is, spin is the mind-pixel (1). According to our theory, the nuclear spins and possibly electron spins inside neural membranes and proteins form various entangled quantum states through action potential modulated nuclear spin interactions and paramagnetic O2/NO driven activations and, in turn, the collective dynamics of the said entangled quantum states produces consciousness and influences the classical neural activities through spin chemistry (1).

As with other quantum mind theories, decoherence is a major concern as pointed out by Tegmark (28) but may not be insurmountable (29). We are convinced that the solution lies with quantum entanglement. Indeed, our dualistic approach adopted earlier allows mind to utilize quantum entanglement to achieve the unity of mind in pre-spacetime (1). The essential question is then how does mind process and harness the information from the mind-pixels which form various entangled states so that it can have conscious experience. We have argued that contextual, irreversible and non-computable means within pre-spacetime are utilized by mind to do this.

6.  **Conclusion**

In this paper we have discussed the ontological origin, implications and potential applications of gravity by thinking outside mainstream notions of general relativity and quantum gravity. We have proposed that gravity originates from the primordial self-referential spin processes in non-spatial and non-temporal pre-spacetime, is the macroscopic manifestation of quantum entanglement, implies instantaneous interconnectedness of all maters in the universe and, once better understood and harnessed, has far-reaching consequences and applications in many areas such as instantaneous communication, spacetime engineering and space travel. We have also discussed the role of gravity in our spin-mediated consciousness theory from this new perspective. Finally, the principle of science dictates that a hypothesis/proposition should only achieve scientific legitimacy if it is experimentally verified. Thus, we have designed and carried out experiments to verify our propositions and the results will be reported separately.

# APPENDIX III

## Thinking Outside the Box III: How Mind Influences Brain Through Proactive Spin

Huping Hu & Maoxin Wu
(Dated : June 8, 2007)

**ABSTRACT**

Benjamin has written an article entitled "Dark Chemistry or Psychic Spin Pixel?" which promotes a "dark chemistry" model of mind and discuss the spin-mediated theory. This hypothetical chemistry is based on the hypothetical axion dark matter. Based on our recent experimental findings, our contentions are two-fold: (1) dark matter is likely the cosmological manifestation of quantum entanglement; and (2) the hypothetical axion dark matter is, therefore, replaceable by non-local effects mediated by the primordial spin processes. We also discuss the cause of apparent dark energy. In particular, we explore the issue how mind influences the brain through said spin processes. Our thoughts are that the manifestation of free will is intrinsically associated with the nuclear and/or electron spin processes inside the varying high electric voltage environment of the neural membranes and proteins which likely enable the said spin processes to be "proactive," that is, being able to utilize non-local energy (potential) and quantum information to influence brain activities through spin chemistry and possibly other chemical/physical processes in defiance of the second law of thermodynamics.

**Key Words**: dark matter, dark energy, axion, gravity, quantum entanglement, proactive spin, free will, consciousness

## 1.    On "Dark Chemistry" Model

Benjamin has written an article entitled "Dark Chemistry or Psychic Spin Pixel?" (Benjamin, 2007) which promotes a "dark chemistry" model of mind and discuss the spin mediated theory (Hu & Wu, 2002, 2004a & 2004b) and the Hameroff-Penrose model (Hameroff & Penrose, 1996). One may recall that dark matter is a hypothetical matter of unknown composition whose presence is inferred from its gravitational effects on visible matter, dark energy is a hypothetical form of energy that permeates all of space and tends to increase the apparent rate of expansion of the universe, and axion is a hypothetical elementary particle postulated to resolve the lack of CP-violation in the physics of quarks and gluons (Source: Wikipedia). Axion is a candidate for dark matter. Benjamin's "dark chemistry" is a hypothetical chemistry based on the hypothetical axion dark matter. Benjamin claims that his model is necessary because "quantum parameters such as spin are universal, while discernible mental phenomena are not[, so] a homunculus seems necessary to provide an adequate ontological substrate for mind, to avoid an integration of infinite regress" (Benjamin, 2007).

Benjamin is commendable for boldly going where no one has gone before. However, we are afraid that Benjamin, possibly a few other individuals, has misunderstood what is spin-mediated consciousness theory. Our theory is an ontological theory and, in its dualistic embodiment, a non-spatial and non-temporal pre-spacetime (nonlocal domain) is the homunculus (Hu & Wu, 2002, 2004a & 2004b). Our concept of spin, being the primordial self-referential process, is much more than a passive quantum parameter. It is the mind-pixel, the linchpin between mind and the brain and the ultimate stop-gap of infinite regress through self-reference in pre-spacetime; or, as Benjamin put it, spin is "psychic" (id). It is through the self-referential spin processes that a conscious being such as a human perceives and interacts with the external/physical world. In other words, spin is a process capable being "proactive" in the brain. To justify such view, we would like to point out that in both Hestenes' geometric formulation of quantum mechanics (Hestenes, 1983) and Bohm's non-local hidden variable formulation it has been shown that spin is solely responsible for all the quantum effects (Esposito, 1999). Further, although spin is universal, the reason why it allows the brain to have conscious experience and free will is because of the particular structures and dynamics of the brain as discussed elsewhere (Hu & Wu, 2002, 2004a & 2004b) and further below.

As we understand it, the gist of Benjamin's model is a homunculus or invisible axion body running along the visible physical body which interacts with said physical body through dark chemistry and serves as the host of the soul/spirit in order to avoid infinite regress. Further, according to Benjamin, the axion body is made of intransient non-electric particles and virtually a hologram integrating the patterns of information at various levels. Presumably (we guess), dark chemistry involves "the resonance between the dark and visible bodies of an organism" and "dynamic biophoton process of kindling and quenching the "potential" of the [holographic] pattern" (Benjamin, 2007).

We agree with Benjamin that "[m]ind and consciousness need not be mystical or magical" (Benjamin, 2007). However, his "dark chemistry" model is very convoluted because it involves all these hypothetical and/or exotic entities such as dark matter, dark particle, dark body, invisible axion, non-electric particle, homunculus, biophoton and graviton. Someone has already commented that "I think that having learned how many new entities are put here into play to explain consciousness, William [of] Occam would turn in his tomb" (Patlavskiy, 2007).

In the following sections, we shall argue that Benjamin may find himself still in the "bright" territory instead of the "dark" side, if he is willing to do an exercise with Occam's razor, cutting out "dark" things and replacing them with non-local effects mediated by the "psychic" spin. Based on our recent experimental findings (Hu & Wu, 2006a-d; 2007a), our contentions are two-fold: (1) dark matter is likely the cosmological manifestation of quantum entanglement; and (2) Benjamin's hypothetical dark matter axion, therefore, is replaceable by non-local effects mediated by the primordial spin processes. We will also discuss the cause of dark energy and touch upon how universe operates without Big Bang and speculate what are Black Holes.

In particular, we shall explore the issue how mind influences the brain through the primordial self-referential spin processes which, we admit, have not addressed in detail previously. Our thoughts are that the manifestation of free will is intrinsically associated with the nuclear and/or electron spin processes inside the varying high electric voltage environment of the neural membranes and proteins which likely enable the said spin processes to be "proactive," that is, being able to utilize non-local energy (potential) and quantum information in pre-spacetime (nonlocal domain) to influence brain activities through spin chemistry and possibly other chemical/physical processes in defiance of the second law of thermodynamics.

## 2. The Origin of Dark Matter and Dark Energy

Before we go on, we would like to state that Einstein is no doubt one of the greatest minds ever lived. He made monumental contributions to physics from explaining photoelectrical effect and Brownian motion to constructing special theory of relativity, the famous formula $E=mc^2$ and Bose-Einstein Statistics. But just because Einstein is great does not mean that he was infallible as the case with EPR (Einstein at al, 1935) debate and his general theory of relativity ("GTR," see Einstein, 1915). We are all human and fallible. In any case, whatever happens, GTR is till an effective (approximate) theory for some parts of the universe such as our own solar system.

With this being said, it is likely, we contend, that Einstein's GTR is ontologically invalid because our experimental results indicate that gravity is nonlocal and instantaneous (Hu & Wu, 2006a-d; 2007a) as Newton reluctantly assumed (Newton, 1999 by Cohen et al) and Mach conjectured (Mach, 1960 by Open Court Pub. Co.) and a few other authors argued (e.g., Pope & Osborne, 1996). Besides, many experiments have shown that quantum entanglement is physically real (e.g., Aspect, 1982; Julsgaard et al, 2001) which implies that Einstein's theories of relativity are in real not imagined conflict with quantum theory. Until now, relativists have been able to hide behind the no-signaling "veil" because of the Eberhard Theorem (Eberhard, 1978). But that "veil" has been pierced and we must deal with reality. We understand that pointing out the real possibility that "the Emperor (Einstein) has no clothes" as far as GTR is concerned will irritate a great number of scientists in the mainstream, especially those on the superstring bandwagon, and may eventually destroy jobs, livelihood and research grants. But we need to ask ourselves the soul searching question: Are we here for truth and the greater benefit of mankind or our self-interests? and do we want to go down in history as conniving hypocrites or truth-seeking scientists? And so, as John F. Kennedy would urge, my fellow Scientists: ask not what mankind can do for you but what can you do for mankind.

We have proposed in a previous paper that: (1) gravity originates from the primordial spin processes in non-spatial and non-temporal pre-spacetime (nonlocal domain) and is the macroscopic manifestation of quantum entanglement; and (2) thus, gravity is nonlocal and instantaneous which implies that all matters in the universe are instantaneously interconnected and many anomalous effects in astronomy such as dark matter, dark matter, red shift and Pioneer effect may be resolved from this perspective (Hu & Wu, 2007b).

Experimentally, we have found that the gravity of water in a detecting reservoir quantum-entangled with water in a remote reservoir can change when the latter was remotely manipulated such that, it is hereby predicted, the gravitational energy/potential is globally conserved (Hu & Wu, 2006a-d; 2007a). We have also found that the pH value and temperature of water in a detecting reservoir quantum-entangled with water in a remote reservoir changes when the latter is manipulated under the condition that the water in the detecting reservoir is able to exchange energy with its local environment (*id*). Thus, among other things we have realized non-local signaling using three different physical observables and experimentally demonstrated Newton's instantaneous gravity and Mach's instantaneous connection conjecture and the relationship between gravity and quantum entanglement. Our findings also imply that the properties of all matters can be affected non-locally through quantum entanglement mediated processes. Second, the second law of thermodynamics may not hold when two quantum-entangled systems together with their respective local environments are considered as two isolated systems and one of them is manipulated. Third, gravity has a non-local aspect associated with quantum entanglement thus can be non-locally manipulated through quantum entanglement mediated processes (*id*). Therefore, our findings support a non-local cosmology (Hu & Wu, 2007b).

In light of these developments, we now ask the question what is the origin of dark matter and dark energy. To stray a bit, we further "naively" ask the question how universe operates if the Big Bang didn't happen. Since modern Big Bang theory and Black Holes are based on Einstein's GTR, there is a good chance that Big Bang didn't happen and apparent Black Holes are not actually Black Holes. There are many technical and general papers written in these areas too numerous to mention.

It is our current view that the universe is regenerative: It probably had no beginning and will have no ending, but is constantly and dynamically regenerated through cosmological processes associated with the primordial self-referential spin processes (Hu & Wu, 2003 & 2004b). These spin processes have two aspects: one aspect is expressive being associated with the concept of differentiation, negative entropy and David Bohm's unfolding (Bohm & Hiley, 1993); the other is regressive being associated with the concept of un-differentiation, entropy and David Bohm's enfolding (*Id*). In our view dark matter is the cosmological manifestation of quantum entanglement associated with the regressive and un-differentiating aspect of the primordial self-referential spin process (Hu & Wu, 2003 & 2004b) but seen as additional gravity caused by invisible matter under some cosmological conditions. In contrast, dark energy is the cosmological manifestation of reverse quantum entanglement associated with the expressive and differentiating aspect of the primordial self-referential spin process but seen as anti-gravity caused by negative pressure on the cosmological scale.

It is also our current view that entropy is really about regression and un-differentiation in which explicate and differentiated orders regress/un-differentiate (or become chaotic or random due to missing information, that is, our own ignorance due to complexity) through spin-mediated enfolding (quantum entanglement). On the other hand, it is our view that negative entropy is really about expression and differentiation in which hidden orders in pre-spacetime become explicate and differentiated under the right conditions through spin-mediated unfolding (reverse quantum entanglement) in which the second law of thermodynamics does not apply.

Besides the ever-evolving Life on earth, where can we find these expressions in a dynamic and regenerative universe? The famous but controversial Russian physicist N.A. Kozyrev suggested long time ago that stars, such as our own Sun, are machines generating energy through "active time" (e.g., Kozyrev, 2006 in PiP) instead of nucleosysthesis which would died out or not exist at all if the universe had no beginning. The energy researcher Harold Aspden has also for a long time advocated the view that the main source of the Sun's energy is not from nucleosynthesis but the age-old ether which fills the vacuum of space and could be converted into thermal radiation because of the particular composition and structure of the Sun (e.g., Aspden, 2006). Well, one may not agree with the details of the Kozyrev or Aspden model, we suggest that the Sun may well be producing thermal radiations through spin-mediated expressive processes converting nonlocal energy (potential) in pre-spacetime into regular energy in spacetime.

Further, there is also the possibility that those apparent Black Holes are actually the centers of more violent regenerative cosmological processes in display. On the one hand, these structures violently express (unfold) nonlocal energy into visible matter with said expressive process being seen as dark energy. But, on the other hand, they also violently crush (enfold) visible matter into nonlocal energy with the patterns of said visible matter being seen as the symptom of a Black Hole and said regressive process being seen as dark matter. Of course, other authors probably have already expressed similar views from different angles or perspectives (See, e.g., Pope & Robinson, 2007).

### 3.   How Mind Influences Brain through "Proactive" Spin

We will now explore how mind influences the brain, through the primordial self-referential spin processes (Hu & Wu, 2002, 2003 & 2004a&b), which, we admit, have not addressed in detail previously. All we have said was that the collective dynamics of nuclear and/or electron spin ensembles in neural membranes and protein is able to affect the neural activities of the brain through spin chemistry (*Id*). However, in order to purposefully influence neural activities inside the brain so that a conscious being such as a human can interact with the external world and have free will, the said nuclear and/or electron spin ensembles have to be able to either self-organize to produce free-will-enabling emergent property which cannot be deduced from the spin properties or self-refer to their primordial origin, the non-spatial and non-temporal pre-spacetime (nonlocal domain) which in a dualistic embodiment hosts the mind and is the container of nonlocal energy (potential).

As we have discussed elsewhere, the brain is an electrically very active place where the electric field strengths inside the neural membranes and proteins during a typical action potential oscillates between -9 to +6 million volts per meter which are comparable to those causing electroporation of cell membranes and dielectric breakdown of many materials (Hu & Wu, 2004c&d). So, these electrical fields and their modulations through the action potentials significantly affect the conformations and orientations of neural membrane components such as phospholipids, cholesterols and proteins. Indeed, voltage-dependent ion channels perform their functions through electric field induced conformation changes of the constituent proteins and studies on the effects of electric fields on lipids support the above conclusion (*Id*). We have shown that nuclear spin networks in neural membranes are modulated by action potentials through J-coupling, dipolar coupling and chemical shielding tensors and perturbed by microscopically strong and fluctuating internal magnetic fields produced largely by paramagnetic oxygen (*Id*). We have suggested that these spin networks could be involved in brain functions since said modulation inputs information carried by the neural spike trains into them, said perturbation activates various dynamics within them and the combination of the two likely produce stochastic resonance thus synchronizing said dynamics to the neural firings (*Id*).

Here we specifically propose the following: (1) the varying high electric voltages, being modulated by the action potentials inside the neural membranes and proteins, not only are able to input information into the nuclear and/or electron spin ensembles inside them but also are able to change the characters and properties of these spin ensembles and the pre-spacetime associated with them, making these spins to be "proactive;" and (2) the "proactive" spins so enabled allow the mind to utilize non-local energy (potential) and quantum information to influence brain activities through spin chemistry and possibly other chemical/physical processes in defiance of the second law of thermodynamics.

What plausible evidence do we have to support our above, some would say "outrageous," proposition? The answer is that we indeed do have some evidence supporting this proposition. First, our own recent experiments as discussed earlier and elsewhere shows that nonlcoal signaling and nonlocal effects mediated by quantum entanglement are physically real (Hu & Wu, 2006a-d; 2007a). Our results also imply that systems driven by quantum information such as our brain may defy second law of thermodynamics (*Id*). Second, the well-known placebo effect clearly indicates the influence of the mind over body.

Further, there are many experimental reports in parapsychology showing the possibility or at least plausibility of mind's influences over brain or matter. Of course, one always needs to be very careful about drawing conclusions and inferences from these reports. Just to mention a few, the PEAR Lab's results accumulated over the years shows that mind could alter, however small, random number sequences (e.g., Jahn & Dunn, 2005). William Tiller has reported that under a particular circumstance mind could influence the pH value of water in a remote location through an embedded device (e.g., Tiller, 2007). Danielle Graham and her group have recently reported anomalous gravitational and electromagnetic effects of certain trained persons during meditations (Graham, 2006). Indeed, Dean Radin has documented many related results in his most recent book and was able to repeat and verify some of these results through the studies of his own group (Radin, 2006).

In addition, in the areas of alternative energy, commonly labeled as the "back water" of energy research by the mainstream, there are numerous reports of excess heat being produced through electrophoreses and various plasma discharge schemes both in water and vacuum tubes (e.g., Graneau & Graneau, 1983; Correa & Correa, 2004). The common feature shared by these reports is that somehow under the influence of electric fields or high electric voltages, excess heat was claimed to be produced from the vacuum or age-old ether. If some of these claims are true, we suggest that the source of the excess heat is the nonlocal energy (potential) in pre-spacetime (nonlocal domain) which under the particular arrangements in those experiments was converted into regular energy in spacetime such as thermal radiation.

## 5.   Conclusion

In this paper we have responded to Benjamin's "dark chemistry" perspective. However, we have argued that his hypothetical dark matter axion is replaceable by non-local effects mediated by the primordial self-referential spin processes. This argument is based on our recent experimental findings which suggest that dark matter is likely the cosmological manifestation of quantum entanglement. In particular, we have explored the issue how mind influences the brain through the primordial self-referential spin processes. Our current thoughts are that the manifestation of free will is intrinsically associated with the nuclear and/or electron spin processes inside the varying high electric voltage environment of the neural membranes and proteins which likely enable the said spin processes to be "proactive," that is, being able to utilize non-local energy (potential) and quantum information to influence brain activities through spin chemistry and possibly other chemical/physical processes in defiance of the second law of thermodynamics. We have also discussed what dark energy is and touched upon the issues of Big Bang and Black Holes and made "naïve" speculations about them. If anyone gets offended by these thoughts, we sincerely apologize.

Finally, we cannot stress enough that "talk is cheap" and what really matter are what can be observed and measured experimentally. So our emphasis is still experimental studies and we have and will continue to "put our money where our mouth is" and let experimental observations and measurements to speak for themselves.

ACKNOWLEDGEMENT: We wish to thank Robert Boyd for bringing to our attention N. A. Kozyrev' work in his recent group e-mail.

# APPENDIX IV

## Concerning Spin as Mind-pixel: How Mind Interacts with Brain through Electric Spin Effects

Huping Hu and Maoxin Wu

(Dated: November 6, 2007)

**Abstract:** Electric spin effects are effects of electric fields on the dynamics/motions of nuclear/electron spins and related phenomena. Since classical brain activities are largely electric, we explore here a model of mind-brain interaction within the framework of spin-mediated consciousness theory in which these effects in the varying high-voltage electric fields inside neural membranes and proteins mediate mind-brain input and output processes. In particulars, we suggest that the input processes in said electric fields are possibly mediated by Dirac-Hestenes electric dipoles and/or spin transverse forces both of which are associated with the nuclear/electronic spin processes. We then suggest that the output processes (proactive spin processes) in said electric fields possibly also involve Dirac-Hestenes electric dipole interactions in said electric fields and Dirac negative energy extraction processes, shown by Solomon, of nuclei/electrons besides non-local processes driven by quantum information. We propose that these output processes modulate the action potentials, thus influencing the brain, by affecting the cross-membrane electric voltages and currents directly and/or indirectly through changing the capacitance, conductance and/or battery in the Hudgkin-Huxley model. These propositions are based on our own experimental findings, further theoretical considerations, and studies reported by others in the fields of spintronics, high-energy physics and alternative energy research.

**Key Words:** spin; mind-pixel; electric spin effect; spin transverse force; Dirac-Hestenes electric dipole; electric field; proactive spin

## 1. Introduction

Within the framework of spin-mediated consciousness theory, the nuclear and/or electronic spins are proposed to be the mind-pixels which interact with the brain through quantum effects, modulating and being modulated by various classical brain activities such as the action potentials (Hu & Wu, 2002 & 2004a-d). Since classical brain activities are largely electric and, in comparison, magnetic fields insides the brain are only microscopically strong but fluctuating, we have previously discussed how action potentials modulate the dynamics of nuclear/electron spin networks inside the brain through J-coupling, dipolar coupling and chemical shielding tensors, thus, feeding information into mind in the dualistic approach (Hu & Wu, 2004 c & 2004d). Further, based on our own experimental findings and work done by others, we have also discussed on how mind might influence brain through proactive spin processes enabled by the varying high-voltage electric fields inside the brain (Hu & Wu, 2006a-d & 2007a-c).

What we have not explored in details so far are the electric spin effects which are the direct effects of electric fields on the dynamics/motions of nuclear and/or electronic spins and related phenomena, and their possible roles in mind-brain interactions.

Therefore, we explore here a more specific model of mind-brain interaction within aforesaid framework in which the said electric spin effects in the varying high-voltage electric fields inside neural membranes and proteins mediate mind-brain input and output processes. We first suggest that the input processes in said electric fields are possibly mediated by Dirac-Hestenes electric dipoles and/or spin transverse forces both of which are associated with the nuclei/electron spin processes. We then suggest that the output processes (proactive spin processes) in said electric fields possibly involve Dirac-Hestenes electric dipole interactions with in said electric fields also and Dirac negative energy extraction processes, as shown by Solomon (2006 &2007), of nuclei and/or electrons besides non-local processes driven by quantum information shown by us. We suggest that these output processes modulate the action potentials, thus influencing the brain, by affecting the cross-membrane electric voltages and currents directly and/or indirectly through changing the capacitance, conductance and/or battery in the Hudgkin-Huxley model. These propositions are based on our own experimental findings, further theoretical considerations within the framework of spin-mediated consciousness theory, and studies reported by others in the fields of spintronics, high-energy physics and alternative energy research.

## 2. Dirac-Hestenes Electric Dipole

It has been long known that in an external electric field, the Dirac particle such as an electron or nuclear sub-entity acts as if it has an imaginary electric moment $|\mathbf{d}|=i\mathrm{eh}/4\pi\mathrm{mc}$. Dirac was aware this in 1928 and wrote that "[t]he electric moment, being a pure imaginary … should not…appear in the model [and it] is doubtful whether the electric moment has any physical meaning ..." (Dirac, 1928). Later, Dirac stated that "these extra terms involve some new physical effects, but since they are not real they do not lend themselves very directly to physical interpretation" (Dirac, 1983).

It was Hestenes who showed that Dirac magnetic and electric dipole moments have same origin associated with spin and magnetization (For a review, see, Hestenes, 2003). In Hestenes' formulism, magnetic moment density is not directly proportional to the spin but "dually proportional." The duality factor $e^{i\beta}$ has the effect of generating an effective electric dipole moment for the Dirac particle. Hestenes commented that "this seems to conflict with experimental evidence that the electron has no detectable electric moment, but the issue is subtle" (*Id*). Other researchers have also shown recently that the magnetic and electric dipole moments of a fermion are closely related because they determine the real and imaginary part of the same physical quantity (Feng *et al*, 2001; Graesser & Thomas, 2002).

Indeed, Silenko has recently shown in the Foldy-Wouthuysen representation that although the influence of the electric dipole moment on the Dirac particle motion is negligibly small in an external electric field, it influences significantly the spin motion of the said particle (Silenko, 2006).

In addition, in the classical models of the Dirac particle, fast oscillating electric dipole moments also appear (Rivas, 2005; Gauthier 2006). These findings coincide with earlier finding that a moving magnetic dipole induces an electric dipole $\mathbf{d}=(\mathbf{v}/c^2)\times\mathbf{m}$, where $\mathbf{m}$ and $\mathbf{v}$ are respectively the magnetic moment and the velocity of the moving spin, as a relativistic effect (Rosser, 1964). Rivas (2005) believes that what is lacking in the typical quantum mechanical wave equation is this oscillating electric dipole. He states that "in general, the average value of this term in an electric field of smooth variation is zero, [but] in high intensity fields or in intergranular areas in which the effective potentials are low, but their gradients could be very high, this average value should not be negligible." Rivas further showed that the electric moment of the classical Dirac electron could lead to interesting physical effects (*Id*).

Here we specifically propose (Proposition I) that in the dualistic mind-brain approach of spin-mediated consciousness theory the interactions between the Dirac-Hestenes electric dipoles of nuclei and/or electrons with the varying high-voltage electric fields inside the neural membranes and proteins directly feed information carried by the neural spike trains into mind through the varying high-voltage action potentials.

Do we have any other justifications for Proposition I besides the points already discussed above? The answer, indeed, is "Yes." First, even if the Dirac electric dipole is purely imaginary with no known physical consequence, we argue that in the dualistic mind-brain approach, it may serve as an information receiver in the non-local domain where mind resides for the simple reason that such non-local domain is likely amicable to a description by the imaginary numbers (See, e.g., Rauscher & Targ, 2001).

Second, following Hestenes (see, *e.g.*, 2003) and possibly others, we strongly believe that the origin of the electric dipole is intrinsically associated with a Dirac particle actually being a composite entity with the un-manifested antiparticle inseparably accompanying the regular particle. Since the antiparticle sometimes shows up in our spacetime as real, we have reason to believe that the same in its un-manifested form is an active participant in the primordial self-referential spin processes driving quantum mechanics, spacetime dynamics and consciousness as will be discussed elsewhere in due time (Also see, Hu & Wu, 2003 & 2004b).

## 3. Spin Transverse Force

Recent studies in spintronics have shown that an electric field will exert a transverse torque/force on a moving spin (see, *e.g.*, Sun *et al* 2004; Shen, 2005). This is actually not hard to understand since according to special theory of relativity a moving spin in an electric field sees a magnetic field.

Sun *et al* (2004) has shown that a moving spin is affected by an external electric field and feels a force/torque as **m**×[(**v**/$c^2$)×**E**] where **m** and **v** are respectively the magnetic moment and the velocity of the moving spin and **E** is the external magnetic field.

Shen (2005) has shown that, as a relativistic quantum mechanical effect, an external electric field exerts a transverse force on an electron spin 1/2 if the electron is moving. The said spin force, analogue to the Lorentz for on an electron charge in a magnetic field, is perpendicular to the electric field and the spin motion when the spin polarization is projected along the electric field (*Id*).

Indeed, while this paper is been written, this effect has just been successfully used in the laboratory to flip the spin of an electron in a quantum dot by applying an oscillating electric field (Nowack, *et al*, 2007). The electric field induces coherent transitions (Rabi oscillations) between spin-up and spin-down with 90° rotations as fast as ~55 ns and the analysis done by the authors indicates that the electrically-induced spin transitions are mediated by the spin-orbit interaction (*Id*).

Therefore, we specifically propose here (Proposition II) that the interactions between the moving nuclear/electronic spins in neural membranes and proteins and the varying high-voltage electric fields there directly feed information into mind in the dualistic mind-brain approach of spin mediated consciousness theory.

To illustrate this particular mechanism, we now consider the spin transverse force exerted on a proton spin of a hydrogen atom connected to the carbon chain of a phosphate lipid located inside the neural membranes as shown in Figure 1. As the carbon chain rotates in parallel to the intense electric field **E** across the neural membranes, the vertical proton spin moving in a circle perpendicular to the carbon chain sees a magnetic field in the rotating frame of reference thus feels a transverse torque/force *f* toward the rotating plane. Quantitative calculations shall be performed in a separate paper.

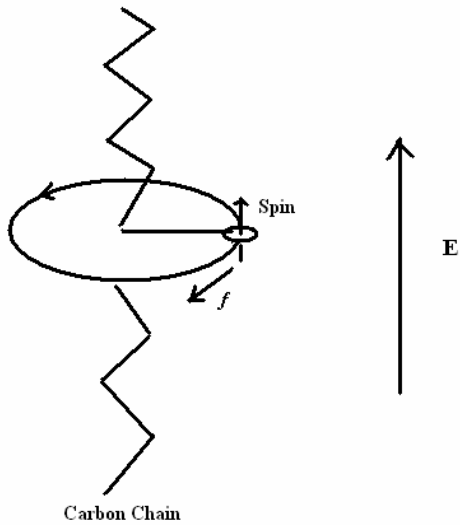

Fig.1. Illustration of spin transverse torque/force *f* exerted on a nuclear/electronic spin on a molecular chain or fragment inside the neural membranes and proteins.

This spin transverse torque/force enables the neural spike trains to directly influence the nuclear/electronic spin networks in neural membranes and proteins thus inputting information into mind in the dualistic approach.

**4. Output Process through Proactive Spin**

Previously, we have explored how mind influences the brain through primordial self-referential spin processes (Hu & Wu, 2007c). Our thoughts were that the manifestation of free will is intrinsically associated with the nuclear and/or electron spin processes inside the varying high- voltage environment of the neural membranes and proteins which likely enable the said spin processes to be proactive, that is, being able to utilize non-local energy (potential) and quantum information to influence brain activities through spin chemistry and possibly other chemical and/or physical processes in defiance of the second law of thermodynamics (*Id*).

With respect to possible evidence, we pointed out that: (1) our own experiments suggest that non-local energy/potential and quantum information may influence brain function through quantum entanglement mediated processes (Hu & Wu, 2006a-d; 2007a); (2) the well-known placebo effect clearly indicates the influence of the mind over body; (3) many experimental reports in parapsychology indicate the plausibility of mind's influences over brain or matter (e.g., Jahn & Dunn, 2005; Tiller, 2007; Graham, 2006; Radin, 2006); and (4) in the field of alternative energy research, there are numerous papers reporting excess heat being produced through electrophoreses and various plasma discharge schemes the common feature of which is that somehow under the influence of electric fields or high electric voltages, excess heat was claimed to be produced from the vacuum or age-old ether (e.g., Graneau & Graneau, 1983; Correa & Correa, 2004).

Here we suggest possible additional mechanisms besides what our own experiments have shown. One such possible additional mechanism is connected to the Dirac-Hestenes electric dipole associated with nuclear/electronic spin. As Rivas (2005) pointed out that what is lacking in the quantum mechanical wave equation is possibly the oscillating electric dipole which in high intensity fields or in intergranular areas should not be negligible in its contributions to dynamics of the Dirac particle.

The complication is that the appearance of an imaginary dipole energy term $H_{int} = (-i\mathbf{d}.\mathbf{E})$ in the Hamiltonian makes it non-hermitian and the corresponding energy spectrum complex-valued instead of real valued. However, various studies of the non-hermitian Hamiltotians indicate that not only very interesting novel dynamics appear due to non-hermiticity (See, *e.g.*, Aguiar Pinto *et al*, 2003) but also certain classes of non-hermitian Hamiltonians still have real energy spectra (See, e.g., Bender, 2007). We propose (Proposition III) that these new dynamics may be just what are needed to enable the proactive spin process.

Another possible mechanism is the Dirac negative energy extraction in a varying electric field which has been shown to be theoretically possible by Solomon (e.g., Solomon, 2006 & 2007). The vacuum electrons obey the Dirac equation and the energy of these electrons will change in response to an applied electric field (*Id.*). Solomon has examined the vacuum in Dirac's hole theory and he showed that it is possible to find a varying electric field for which the change in the energy of each vacuum electron is negative (*Id.*). Therefore, according to Solomon, the total change in the energy of the vacuum state is negative and this new state will have less energy than the original unperturbed vacuum state (*id*).

Solomon's theoretical results provide support to the claims of excess heat being produced through electrophoreses and various plasma discharge schemes (e.g., Graneau & Graneau, 1983; Correa & Correa, 2004).

We suspect that the Dirac negative energy extraction process shown by Solomon is connected to the dynamics of the Dirac-Hestenes electric dipole in a varying electric field. Further, we don't think that a vacuum electron in a varying electric field can fall into the negative energy state occupied by the un-manifested antiparticle.

Instead, we propose (Proposition IV) that in certain varying external electric field the vacuum electron pumps energy from the vacuum and release the same in our spacetime in order to maintain its minimal energy state in the vacuum. We will discuss where the vacuum energy comes from in a separate paper.

We further propose (Proposition V) that these output processes modulate the action potentials, thus influencing the brain, by affecting the cross-membrane electric voltages and currents directly and/or indirectly through changing the capacitance, conductance and/or battery in the Hudgkin-Huxley model.

## 5. Conclusion

Electric spin effects are effects of electric fields on the dynamics of nuclear and/or electronic spins and related phenomena. In this paper, we have explored several such effects and their possible roles in the mind-brain interactions within the framework of spin mediated consciousness theory. We have outlined five propositions. In particulars, we have considered a more specific model of mind-brain interaction in which these effects in the varying high-voltage electric fields inside neural membranes and proteins mediate mind-brain input and output processes.

We have suggested that the input processes in said electric fields are possibly mediated by Dirac-Hestenes electric dipoles (Proposition I) and/or spin transverse force/torque (Proposition II) both of which are associated with the nuclear/electronic spin processes. We then suggest that the output processes (proactive spin processes) in said electric fields possibly also involve Dirac-Hestenes electric dipole interactions in said electric fields (Proposition III) and Dirac negative energy extraction processes, as shown by Solomon, of nuclei/electrons (Proposition IV) besides non-local processes driven by quantum information. We have proposed (Proposition V) that these output processes modulate the action potentials, thus influencing the brain, by affecting the cross-membrane electric voltages and currents directly and/or indirectly through changing the capacitance, conductance and/or battery in the Hudgkin-Huxley model.

These propositions are based on our own experimental findings, further theoretical considerations within the framework of spin-mediated consciousness theory, and studies reported by others in the fields of spintronics, high-energy physics and alternative energy research.

Finally, we stress again the importance of experimental work and quantitative calculations and computer simulations in order to verify these propositions and make substantial progresses. We have done some preliminary work in these directions. As usual, results will be reported, when feasible.